%% file: ms_arxiv.tex
\documentclass[twoside,12pt,a4paper]{article}
\usepackage[top=25mm,bottom=25mm,left=25mm,right=25mm]{geometry}
\usepackage{hyperref,nameref,relsize,float,graphicx,import,color,pgfplots,tikz,pgfplotstable}
\usetikzlibrary{arrows,pgfplots.groupplots,patterns}
\pgfplotsset{compat=newest,tick style={black,semithick} }
\usepackage{amsmath,amsfonts,amssymb}
\usepackage{setspace}
\usepackage[utf8]{inputenc}
\usepackage{authblk}
\usepackage{astron}
\pdfoutput=1
\title{Stability of traveling, pre-tensioned, heavy cables}
\date{}
\author{Abhinav R. Dehadrai\thanks{Email: abhinavd@iitk.ac.in} }
\author{Ishan Sharma}
\author{Shakti S. Gupta}
\affil{Mechanics \& Applied Mathematics Group, Department of Mechanical Engineering, Indian Institute of Technology Kanpur, Kanpur 208016, India}

\begin{document}
\maketitle
\allowdisplaybreaks
\begin{abstract}
We study the dynamics of an inclined tensioned, heavy cable 
traveling with a constant speed in the vertical plane. The cable is 
modeled as a beam resisting bending and shear. The governing equation for 
the transverse in-plane vibrations of the cable are derived through the 
Newton-Euler method. The cable dynamics is also studied in the limit of 
zero bending stiffness. In all cases, application of energy balance 
reveals that the total energy of the system fluctuates even though the 
oscillations are small and bounded in time, indicating that the system is 
nonconservative. A comprehensive stability analysis is carried out in the 
parameter space of inclination, traveling speed, pre-tension, bending 
rigidity and the slenderness of the cable. Effect of damping is also 
considered. We conclude that, while pre-tension, rigidity and slenderness 
enhance the stability of the traveling cable, the angle of inclination 
affects the stability adversely. These results may act as guidelines for 
safer design and operation.
\end{abstract}
\section{Introduction} \label{sec-intro}
Traveling cables are fundamental driving mechanisms in elevators, conveyor belts, automobile chain-drives, cableways, etc. In these, all or part of the cable is inclined to gravity. 
During operations the cable tends to oscillate transversely as it travels longitudinally \cite{sack1954transverse}, \cite{miranker1960wave}, \cite{swope1963vibrations}, \cite{mote1965study}, \cite{mote1966nonlinear}, \cite{barakat1968transverse},  \cite{thurman1969free}, \cite{rogge1972equations}, \cite{wickert1989energetics}, \cite{wickert1990classical}, \cite{wickert1992non}. 
To prevent breakdown due to fatigue these oscillations need to remain within some design limits.  
In these studies, cables are modeled either as an elastic beam that resists bending moment and shear, or as a string which supports no bending or shear. 

The dynamics of \textit{horizontally} traveling (traveling direction normal to gravity) beams and strings is well studied \cite{sack1954transverse}, \cite{miranker1960wave}, \cite{swope1963vibrations}, \cite{mote1965study}, \cite{mote1966nonlinear}, \cite{barakat1968transverse},  \cite{thurman1969free}, \cite{rogge1972equations}, \cite{wickert1989energetics}, \cite{wickert1990classical}, \cite{wickert1992non}. Most work has focused on modal analysis to obtain natural frequencies (eigenvalues) at different operating speeds. Eigenvalues appear as complex conjugates, in which the imaginary part contributes to oscillations while the real part controls the amplitude. During stable operations the real part is absent. Therefore, the \textit{critical speed} at which the eigenvalue first appears as a conjugate pair is of special interest. 
At this speed the amplitude may grow with time by virtue of the positive real part, leading to instability. 
In the case of horizontally traveling strings, \cite{miranker1960wave} investigated the continuous energy influx from the material outside the boundary. Response to harmonic support excitation was also studied in \cite{miranker1960wave}, and the instability was identified with the resonance frequency, as the amplitude of the vibration grew with time. 
Discussions on dissipation and stability of general dynamical systems are found in \cite{ziegler2013principles}. 

The critical speed of a traveling cable varies with the type of supports and the distance between them. 
The arc-length based model of a spatially traveling string proposed by \cite{rogge1972equations} generalized all its preceding models, and the static equilibrium equation of a string supported at its ends \cite{alekseev1964equilibrium} was obtained as a special case. 
The calculations in \cite{barakat1968transverse} of the energy flow in a dissipation-free model were later improved when an additional energy influx from the supports was identified by  \cite{wickert1989energetics}. However, it was noted that the total energy during steady horizontal travel was not constant in spite of zero damping. Thus, the system was nonconservative whether or not the modal analysis predicted instability. 

Most results on horizontally traveling beams and strings have been compiled by \cite{hagedorn2007vibrations} and \cite{banichuk2013mechanics}. They include discussions on modal, transient and energy analyses. 
We restrict ourselves from visiting them individually, due to constraint of space, and focus primarily on  cables traveling in a vertical plane inclined to gravity. By virtue of this inclination a portion of the  cable's weight is distributed along its length. 
This, in turn, affects the critical speed at which instability sets in. To the best of our knowledge this case has not yet been investigated.

The paper is organized as follows. We begin by modeling the cable as a beam that resists bending and shear, and which travels at an inclination, while resting upon small rollers; see Fig.\,\ref{fig:beam}. A string model, useful for very flexible cables, is obtained as a special case. 
We then perform a modal analysis to identify the relation between speed, inclination, tension, bending rigidity, slenderness ratio and the conditions of instability. Finally, we verify the instability from  evolution of both the energy and the amplitude of oscillations obtained after direct time-integration of the model. The concluding section summarizes the implications on cable design of our analyses. 

\section{Governing equations} \label{sec-goveqn}
\begin{figure}[h!]
\vspace*{2pt}
\begin{center}
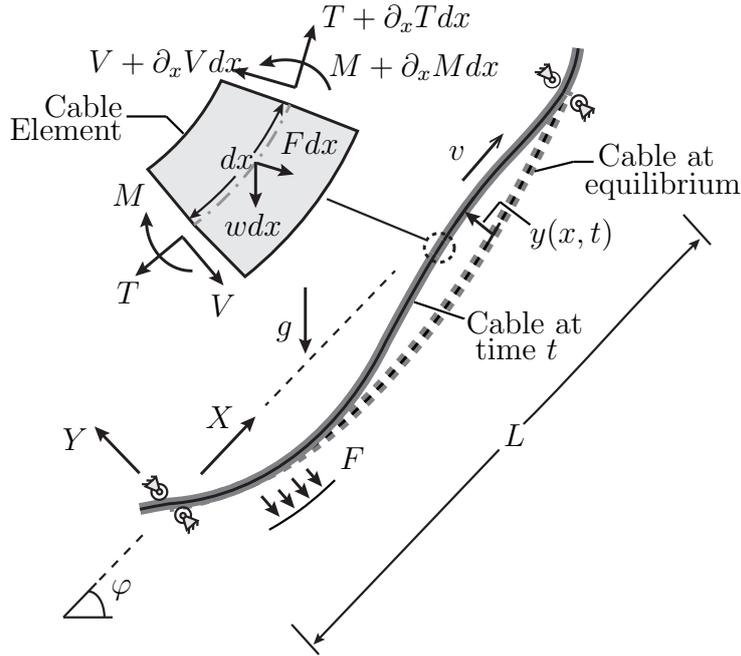
\end{center}
\caption{Schematic of a traveling cable at time $t$ (solid curve) and at its static equilibrium (broken curve). The free-body diagram of a material element is also shown. The deflections of the cable are exaggerated for clarity.}
\label{fig:beam}
\end{figure}

A heavy cable traveling between two pairs of small rollers is shown in Fig.\,\ref{fig:beam}. The cable is inclined at an angle $\varphi$ from the horizontal. The plane is defined using the fixed coordinate system $X$-$Y$, which is aligned with the cable, and has its origin in between the bottom rollers. 
The distance between the top and bottom rollers is $L$ and the acceleration due to gravity $g$ acts vertically downwards. The cable has uniform mass density $\rho$, cross-section $A_c$, Young's modulus $E$, and cross-sectional area moment of inertia $I$. The speed $v$ of each material particle along the length of the cable is constant. 
The displacement $y=y(x,t)$ along the transverse ($Y$-axis) of a material point located at $x$ is measured from its static equilibrium $y_\text{st} (x)$. Both $y$ and $y_\text{st}$ are assumed to be small. 

Employing the notation $\partial^n_x = (\partial^n /\partial x^n)$ and $\partial^n_t = (\partial^n /\partial t^n)$ for the $n^\text{th}$ order partial derivatives with respect to $x$ and $t$, respectively, the total transverse velocity is 
\[
\dot{y} = \frac{d y}{d t} = \partial_t y + v\,\partial_x y,
\]
where $\partial_t y$ is called the \textit{local} velocity and $v\,\partial_x y $ is the \textit{convective} velocity of a material point. The total acceleration is 
\[
\ddot{y} = \frac{d \dot{y}}{d t} = \partial^2_t y + 2 v\,\partial_x \partial_t y + v^2~\partial^2_x y,
\] 
where $\partial^2_t y $ is the \textit{local} acceleration, $2 v\,\partial_x \partial_t y$ is a gyroscopic term 
and $v^2~\partial^2_x y$ is the {convective}  acceleration.
%
\subparagraph*{ }
The free body diagram of an infinitesimal cable element with length $dx$ and weight per unit length $w = \rho A_c g $ is also shown in Fig.\,\ref{fig:beam}. 
To account for possible presence of fluid drag, a transversely distributed loading $F(x,t) = \zeta\,\dot{y}$, with  damping coefficient $\zeta$, is applied on the element. 
The balance of 
angular momentum 
about the centroid of the element yields the shear force 
\[
V = \partial_x (\rho I \ddot{y} - M), 
\]
where $M = EI~ \partial^2_x y$ is the bending moment. For small transverse deflection, the weight of the cable causes its tension to vary 
along the cable's length: 
\[
T(x) = T(L) - \left( L - x \right) w \sin\varphi,
\] 
where we assume that the tension $T(L)$ at $x=L$ is known. 
The linear momentum balance in longitudinal and transverse directions shows that  the equilibrium deflection of the beam is governed by
\begin{subequations}
\label{eq:goveqnstatics} 
\begin{flalign}
&&& \hspace{-2cm} [ E I\partial^4_x - \partial_x \{ T(x) ~\partial_x \} ] y_\text{st} + w \cos\varphi= 0 
&& 
\\
\text{with} 
&&& \hspace{-2cm} y_\text{st}(0) = y_\text{st}(L) =  \partial^2_x y_\text{st}(0) = \partial^2_x y_\text{st}(L) = 0. 
\label{eq:bcstatics} &&
\end{flalign}
\end{subequations}
Similarly, the transverse vibrations of the beam about the equilibrium are governed by
\begin{subequations}
\label{eq:goveqn} 
\begin{flalign}
&&& \hspace{-2cm} [ E I\partial^4_x - \partial_x \{ T(x) ~\partial_x \} ] y
+ \zeta \dot{y} 
 + ( \rho A_c - \rho I \partial^2_x ) \ddot{y} = 0  
&& 
\\
\text{with} 
&&& \hspace{-1.5cm} y(0,t) = y(L,t) = \partial^2_x y(0,t) = \partial^2_x y(L,t) = 0.  
\label{eq:bc} &&
\end{flalign}
\end{subequations}
In \eqref{eq:goveqn}, the term $EI\partial^4_x y $  accounts for bending rigidity, while $\rho I \partial^2_x \ddot{y}$ is the rotary inertia. When $\varphi = 0$, we recover transverse vibrations of a horizontally traveling Rayleigh beam, while the Euler-Bernoulli beam of \cite{hagedorn2007vibrations} is recovered when rotary inertia is also ignored. 
\subsection{Special case: string model}
\label{sec-string}
For a heavy cable which is highly flexible and has negligible shear and bending stiffness, a simple model, resembling the conventional string model, is arrived at by ignoring $EI$ in (\ref{eq:goveqnstatics}) and (\ref{eq:goveqn}). This yields the governing equation for the equilibrium state:
%
\begin{subequations}
\label{eq:goveqnstringstatics}
\begin{flalign}
&&& \hspace{-3cm} [ - \partial_x \{ T(x) ~\partial_x \} ] y_\text{st} + w \cos\varphi= 0    
 &&
\\
\text{with} 
&&& \hspace{-2cm} y_\text{st}(0) = y_\text{st}(L) = 0, \label{eq:bcstringstatics} &&
\end{flalign}
\end{subequations}
%
while transverse vibrations are governed by
\begin{subequations}
\label{eq:goveqnstring}
\begin{flalign}
&&& \hspace{-3cm}[ - \partial_x \{ T(x) ~\partial_x \} ] y
+ \zeta \dot{y} 
 + \rho A_c \ddot{y} = 0  
 &&
\\ 
\text{with} 
&&& \hspace{-2cm} y(0,t) = y(L,t) = 0.  \label{eq:bcstring} &&
\end{flalign}
\end{subequations} 
The equation for a horizontally traveling string considered in  \cite{hagedorn2007vibrations} is retrieved from (\ref{eq:goveqnstring}) by setting $\varphi=0$. 

%

\subsection{Nondimensionalization} \label{subsec-nondim}
For further analysis, the dimensionless forms of the governing equations (\ref{eq:goveqnstatics}) and (\ref{eq:goveqn}) are derived here. To this end, we introduce nondimensional quantities 
\begin{gather*}
\bar{x} = x/L, \bar{y}=y/L, \bar{t} = t (EI/\rho A_c L^4)^{1/2}, \bar{v} = v (EI/\rho A_c L^2)^{-1/2}, 
\\
\mu = \left\lbrace T(L) L^2 /EI \right\rbrace^{1/2}, \varrho = w L^3 / EI, \lambda = ( I/L^2 A_c )^{-1/2}  \text{ and } c = \zeta (EI \rho A_c /L^2)^{-1/2}.
\end{gather*} 
The dimensionless end tension $\mu$ will be employed as a control parameter to vary tension in the cable. With these \eqref{eq:goveqnstatics} becomes
\begin{subequations}
\label{eq:nondimstatics}
\begin{flalign}
&&& \quad [ \bar{\partial^4_x} - \bar{\partial_x} \{ \bar{T}(\bar{x}) ~\bar{\partial_x} \} ] \bar{y}_\text{st}
+ \varrho \cos\varphi = 0
 &&
\\
\text{and} 
&&& \bar{y}_\text{st}(0) = \bar{y}_\text{st}(1) = 
\bar{\partial^2_x}\bar{y}_\text{st}\,(0) = \bar{\partial^2_x}\bar{y}_\text{st}\,(1) = 0,
&&
\end{flalign}
\end{subequations}
while \eqref{eq:goveqn} transforms to
\begin{subequations}
\label{eq:nondim} 
\begin{flalign}
            &&& [ \bar{\partial^4_x} - \bar{\partial_x} \{ \bar{T}(\bar{x}) ~\bar{\partial_x}   \} ] \bar{y}
+ c \dot{\bar{y}} 
+ ( 1 - \lambda^{-2} \bar{\partial^2_x} ) \ddot{\bar{y}} = 0
 && 
\\
\text{and} 
&&& \bar{y}(0,\bar{t}) = \bar{y}(1,\bar{t}) = 
\bar{\partial^2_x}\bar{y}\,(0,\bar{t}) = \bar{\partial^2_x}\bar{y}\,(1,\bar{t}) = 0,
\label{eq:bcnondim}  &&
\end{flalign}
\end{subequations}
%
where $\bar{\partial_x} = \partial/\partial \bar{x}$ and $\bar{\partial_t} = \partial/\partial \bar{t}$, and the nondimensional tension 
\begin{equation}
\bar{T}(\bar{x}) 
= \mu^2 - \left( 1 - \bar{x} \right) \varrho \sin\varphi .
\label{eq:Tnondim}
\end{equation}
For the string of Sec. \ref{sec-string} we employ the following nondimensionalization: 
\begin{gather*}
\bar{x} = x/L, \bar{y}=y/L, \bar{t} = t \left\lbrace T(L)/\rho A_c L^2 \right\rbrace^{1/2}, \bar{v} = v \left\lbrace T(L)/\rho A_c \right\rbrace^{-1/2}, 
\\
\varrho = w L / T(L), \bar{T}(\bar{x}) = T(x)/T(L) \text{ and } c = \zeta L \left\lbrace T(L)/ \rho A_c \right\rbrace^{1/2} 
\end{gather*} 
With these (\ref{eq:goveqnstringstatics}) becomes 
\begin{subequations}
\label{eq:nondimstringstatics}
\begin{flalign}
&&& [ - \bar{\partial_x} \{ \bar{T}(\bar{x}) ~\bar{\partial_x} \} ] \bar{y}_\text{st} + \varrho\cos\varphi = 0,  && 
\\
\text{and}
&&& \qquad \qquad \bar{y}_\text{st}(0) = \bar{y}_\text{st}(1) = 0, \label{eq:bcstringnondimstatics} 
\end{flalign}
\end{subequations}
while (\ref{eq:goveqnstring}) becomes
\begin{subequations}
\label{eq:nondimstring}
\begin{flalign}
&&& \; [ - \bar{\partial_x} \{ \bar{T}(\bar{x}) ~\bar{\partial_x} \} ] \bar{y}
+ c \dot{\bar{y}} 
+ \ddot{\bar{y}} = 0,  &&
\\
\text{and}
&&& \qquad \qquad \bar{y}(0,\bar{t}) = \bar{y}(1,\bar{t}) = 0.  \label{eq:bcstringnondim} &&
 \end{flalign}
\end{subequations}
We now solve for the equilibrium shapes before proceeding to the dynamics of cables.
%

%

\section{Equilibrium shapes} \label{subsec-eqshape}
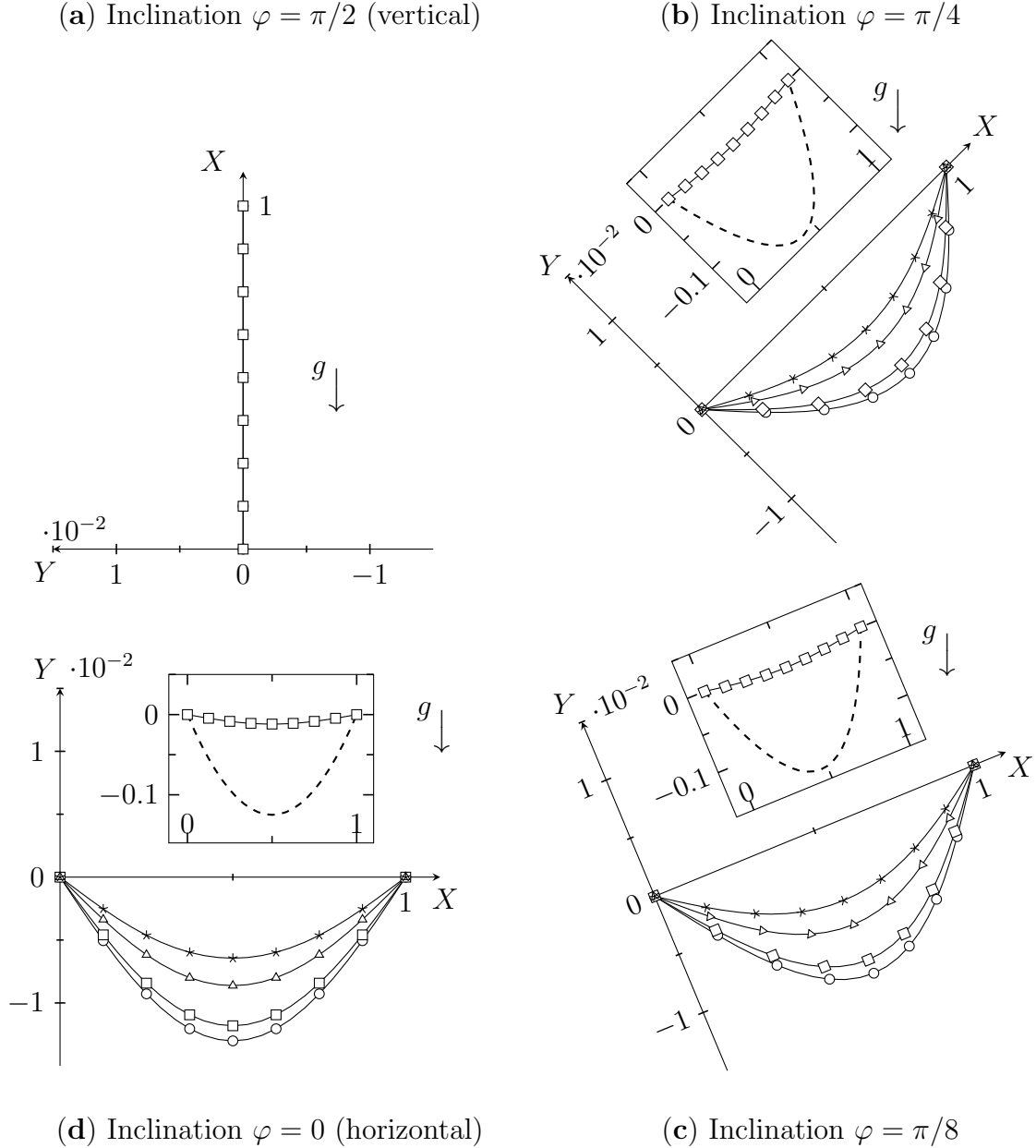
\begin{figure}[h!]
  \begin{center}
%
\pgfplotsset{ 
  every axis plot post/.append style={
   mark repeat=3,
   every mark/.append style={scale=1,fill=white},
   }}

\begin{tikzpicture}
    \begin{axis}
    [
    cycle list name=mark list,
    name=plot1,height=7cm,width=7cm,
    rotate around={90:(current axis.origin)}, 
    domain=0:1,
    x tick label style={rotate=0,anchor=west,},
    y tick label style={rotate=0,anchor=north},
    axis x line=center,
    axis y line=left,
    ymin=-0.015, ymax=0.015,
    xmax=1.1,
    minor tick num=1,xtick={0,1},
    ytick={-0.01,0,0.01},
    ]
    \addplot {0*x}; 
    \addplot { 0*x}; 
    %
     \end{axis}
    \begin{axis}
    [
    cycle list name=mark list,
    name=plot2,at={($(plot1.east)+(1cm,0)$)},anchor=west,height=7cm,width=7cm,
    anchor=origin, 
    xshift=1.5cm,
    yshift=-2.5cm,
    rotate around={45:(current axis.origin)}, 
    domain=0:1,
    x tick label style={rotate=45,anchor=southwest,below},
     y tick label style={rotate=45,anchor=east},
    axis x line=center,
    axis y line=left,
     ymin=-0.015, ymax=0.015,
    xmax=1.1,
    minor tick num=1,xtick={0,1},
    ytick={-.01,0,0.01},
    ]
   \addplot {-0.9586686240e-2*sin(3.141592654*deg(x))-0.1972386300e-4*sin(6.283185308*deg(x))-0.3820966478e-4*sin(9.424777962*deg(x)) }; 
    \addplot { -0.8674992344e-2*sin(3.141592654*deg(x))-0.1742115893e-4*sin(6.283185308*deg(x))-0.3778005493e-4*sin(9.424777962*deg(x)) }; 
    \addplot { -0.6284407732e-2*sin(3.141592654*deg(x))-0.1152083555e-4*sin(6.283185308*deg(x))-0.3615705590e-4*sin(9.424777962*deg(x)) }; 
    \addplot { -0.4674284581e-2*sin(3.141592654*deg(x))-0.7730698372e-5*sin(6.283185308*deg(x))-0.3431734319e-4*sin(9.424777962*deg(x)) };  
    %
    %
    \end{axis}
    %
    %
    \begin{axis}
    [
    cycle list name=mark list,
    name=plot2s,at={($(plot2.north)+(0,1cm)$)},anchor=north,height=4cm,width=4.5cm,
    xshift=-3.1cm,
    yshift=0.3cm,
    rotate around={45:(current axis.origin)}, 
    domain=0:1,
    x tick label style={rotate=45,anchor=southwest,,above},xticklabel pos=left,
     y tick label style={rotate=45,anchor=east},
    ymin=-0.16, ymax=0.05,
    xmax=1.1,
    minor tick num=1,xtick={0,1},
    ytick={-.1,0,0.1},
    ]
    \addplot[mark=square*,mark repeat=3] { -0.8674992344e-2*sin(3.141592654*deg(x))-0.1742115893e-4*sin(6.283185308*deg(x))-0.3778005493e-4*sin(9.424777962*deg(x)) }; 
   \addplot[mark=none,dashed,thick] { .9999999999*x-.7177618087-.8143672782*ln(.414213562+x) }; 
   %
    \end{axis}
    %
    %
    \begin{axis}
    [
    cycle list name=mark list,
    name=plot3,at={($(plot1.south)-(0,1cm)$)},anchor=north,height=7cm,width=7cm,
    rotate around={0:(current axis.origin)}, 
    xshift=0.1cm,
    yshift=-1cm,
    domain=0:1,
    x tick label style={rotate=0,anchor=southwest,below},
     y tick label style={rotate=0,anchor=east},
    axis x line=center,
    axis y line=left,
     ymin=-0.015, ymax=0.015,
    xmax=1.1,
    minor tick num=1,xtick={0,1},
    ytick={-0.01,0,0.01},
    ]
    \addplot { -0.1307105456e-1*sin(3.141592654*deg(x))-0.5379034800e-4*sin(9.424777962*deg(x))}; 
    \addplot { -0.1186852188e-1*sin(3.141592654*deg(x))-0.5319152263e-4*sin(9.424777962*deg(x))}; 
    \addplot {  -0.8675828504e-2*sin(3.141592654*deg(x))-0.5092386662e-4*sin(9.424777962*deg(x))}; 
   \addplot {-0.6492637452e-2*sin(3.141592654*deg(x))-0.4834743733e-4*sin(9.424777962*deg(x)) }; 
    %
    %
    \end{axis}
    %
    %
    \begin{axis}
    [
    cycle list name=mark list,
    name=plot3s,at={($(plot3.north)-(0,1cm)$)},anchor=north,height=4cm,width=4.5cm,
    rotate around={0:(current axis.origin)}, 
    xshift=0.3cm,
    yshift=1.2cm,
    domain=0:1,
    x tick label style={rotate=0,anchor=southwest,,above},xticklabel pos=left,
     y tick label style={rotate=0,anchor=east},
    ymin=-0.16, ymax=0.05,
    xmax=1.1,
    minor tick num=1,xtick={0,1},
    ytick={-0.1,0,0.1},
    ]
    \addplot[mark=square*,mark repeat=3]  { -0.1186852188e-1*sin(3.141592654*deg(x))-0.5319152263e-4*sin(9.424777962*deg(x))}; 
    \addplot[mark=none,dashed,thick]  { -.5*x+.5*x^2};
    %
    \end{axis}
    %
    %
    \begin{axis}
    [
    cycle list name=mark list,
    name=plot4,at={($(plot3.east)-(0,1cm)$)},anchor=north,height=7cm,width=7cm,
    anchor=origin, 
    xshift=3.1cm,
    yshift=-0.5cm,
    rotate around={22.5:(current axis.origin)}, 
    domain=0:1,
    x tick label style={rotate=22.5,anchor=southwest,below},
     y tick label style={rotate=22.5,anchor=east},
    axis x line=center,
    axis y line=left,
     ymin=-0.015, ymax=0.015,
    xmax=1.1,
    minor tick num=1,xtick={0,1},
    ytick={-0.01,0,0.01},
    ]
    \addplot { -0.1231507015e-1*sin(3.141592654*deg(x))-0.1365807068e-4*sin(6.283185308*deg(x))-0.4981136787e-4*sin(9.424777962*deg(x)) }; 
    \addplot { -0.1116176069e-1*sin(3.141592654*deg(x))-0.1208405554e-4*sin(6.283185308*deg(x))-0.4925471095e-4*sin(9.424777962*deg(x)) }; 
    \addplot { -0.8120001379e-2*sin(3.141592654*deg(x))-0.8027870144e-5*sin(6.283185308*deg(x))-0.4714826846e-4*sin(9.424777962*deg(x))  }; 
    \addplot { -0.6056787795e-2*sin(3.141592654*deg(x))-0.5404132791e-5*sin(6.283185308*deg(x))-0.4475680323e-4*sin(9.424777962*deg(x)) }; 
    %
    %
    \end{axis}
    %
    \begin{axis}
    [
    name=plot4s,at={($(plot4.north)-(0,1cm)$)},anchor=north,height=4cm,width=4.5cm,
    anchor=origin, 
    xshift=-1.9cm,
    yshift=0.1cm,
    rotate around={22.5:(current axis.origin)}, 
    domain=0:1,
    x tick label style={rotate=22.5,anchor=southwest,above},xticklabel pos=left,
     y tick label style={rotate=22.5,anchor=east},
    ymin=-0.16, ymax=0.05,
    xmax=1.1,
    minor tick num=1,xtick={0,1},
    ytick={-0.1,0},
    ]
    \addplot[mark=square*,mark repeat=3] { -0.1116176069e-1*sin(3.141592654*deg(x))-0.1208405554e-4*sin(6.283185308*deg(x))-0.4925471095e-4*sin(9.424777962*deg(x)) }; 
    \addplot[mark=none,dashed,thick] {  2.414213562*(x)-5.004865538*ln(1+.6199144049*(x)) }; 
    %
    \end{axis}
    %
    \node at (8,0.1) [black,fill=none,above=0,rotate=-90] {$\longrightarrow$};
    \node at (8,0.1) [black,fill=none,above=0,rotate=0] {$g$};
    \node at (15.2,1.2) [black,fill=none,above=0,rotate=-90] {$\longrightarrow$};
    \node at (15.2,1.2) [black,fill=none,above=0,rotate=0] {$g$};
    \node at (14.5,9) [black,fill=none,above=0,rotate=-90] {$\longrightarrow$};
    \node at (14.5,9) [black,fill=none,above=0,rotate=0] {$g$};
    \node at (6.5,5) [black,fill=none,above=0,rotate=-90] {$\longrightarrow$};
    \node at (6.5,5) [black,fill=none,above=0,rotate=0] {$g$};
    \node at (8.3,-2.6) [black,fill=none,above=0,rotate=0] {$X$};
    \node at (2.6,0.7) [black,fill=none,above=0,rotate=0] {$Y$};
    \node at (16.5,-0.7) [black,fill=none,above=0,rotate=0] {$X$};
    \node at (10,0.2) [black,fill=none,above=0,rotate=0] {$Y$};
    \node at (16,8.5) [black,fill=none,above=0,rotate=0] {$X$};
    \node at (9.8,6.5) [black,fill=none,above=0,rotate=0] {$Y$};
    \node at (5,8) [black,fill=none,above=0,rotate=0] {$X$};
    \node at (2.6,2.1) [black,fill=none,above=0,rotate=0] {$Y$};
    \node at (5.8,-6) [black,fill=none,above=0,rotate=0] {(\textbf{d}) Inclination $\varphi = 0$ (horizontal)};
    \node at (13.5,-6) [black,fill=none,above=0,rotate=0] {(\textbf{c}) Inclination $\varphi = \pi/8$};
    \node at (13.5,10) [black,fill=none,above=0,rotate=0] {(\textbf{b}) Inclination $\varphi = \pi/4$};
    \node at (5.8,10) [black,fill=none,above=0,rotate=0] {(\textbf{a}) Inclination $\varphi = \pi/2$ (vertical)};
 \end{tikzpicture}
%
    \caption{Equilibrium shapes of beams at several  inclinations $\varphi$ held by various dimensionless tensions: $ \mu = 0 $ (--$ \circ $--),  $ \mu = 1 $ (--$ \square $--), $ \mu = 5 $ (--$ \triangle $--) and $ \mu = 10 $ (--$ \ast $--). The equilibrium shapes of a string ($---$) and a beam (--$ \square $--) with tension $\mu=1$ are compared in the insets. The dimensionless mass per unit length $\varrho=1$. } \label{fig:staticeq}
  \end{center}
\end{figure}

The equilibrium shape of an inclined, heavy beam is  found by solving (\ref{eq:nondimstatics}) numerically. The computational algorithm is explained in the next section. The equilibrium shape of a heavy, inclined string, on the other hand, is found from (\ref{eq:nondimstringstatics}) in closed form: 
\[ \bar{y}_\text{st}(\bar{x}) =  - \cot\varphi\,\left\lbrace {G}(\bar{x}) - \bar{x} \right\rbrace\]
{where}
\[{G}(\bar{x}) = \dfrac{ \log \left| { \bar{T}(\bar{x}) }  /  { \bar{T}(0) } \right| }{ \log \left| { \bar{T}(1) }  /  { \bar{T}(0) } \right| } \]
is the percentage gain in tension at location $x$  along the cable's length relative to the tension at the top end. 
When $\varphi = \pi/2$ the string hangs vertically down, while for the horizontal string ($\varphi=0$) the static equilibrium shape 
$\bar{y}_\text{st} = \varrho (\bar{x}^2 - \bar{x}) / 2$ is a parabola\footnote{In general, a simply supported string deforms to the classical catenary (\textit{cosh}) curve under the action of its own weight. However, when the displacement is small, as assumed in Sec. \ref{sec-goveqn}, this curve becomes a parabola.}. The equilibrium shapes of strings and beams at various inclinations and end-tensions are shown in Fig.\,\ref{fig:staticeq}. 
%
%
%

We find from Fig.\,\ref{fig:staticeq} that lower the inclination, closer is the equilibrium shape to a parabola, and lesser is the lengthwise gain in tension. In the inset we compare the string's equilibrium shape with that of a beam
, which carries the same tension $\bar{T}(\bar{x}) =1 - (1-\bar{x})\sin\varphi$ as that of the string. As expected, the string sags more than the beam due to the absence of bending rigidity. 
Finally, the beam bends less with increase in tension.

We next describe the numerical techniques employed for the modal, transient and energy analyses.

\section{Numerical solution} \label{sec-numsol}
%
\pgfplotstableread[col sep = comma]{Vc_phi_lam_EB.txt}\VcphilamEB
\pgfplotstableread[col sep = comma]{Vc_phi_lam_T.txt}\VcphilamT
\pgfplotstableread[col sep = comma]{Vc_phi_lam_EB_T.txt}\VcphilamEBT
\pgfplotstableread[col sep = comma]{Vc_phi_mu_T.txt}\VcphimuT 
\pgfplotstableread[col sep = comma]{Vc_phi_cl_EB.txt}\VcphiclEB
\pgfplotstableread[col sep = comma]{Vc_phi_cl.txt}\Vcphicl
\pgfplotstableread[col sep = comma]{Vc_phi_rho.txt}\Vcphirho
\pgfplotstableread[col sep = comma]{Vc_phi_rho_T.txt}\VcphirhoT
%
%
%
The governing equations (\ref{eq:goveqn}) and (\ref{eq:goveqnstring}), or their dimensionless forms  (\ref{eq:nondim}) and (\ref{eq:nondimstring}), are not self-adjoint. 
The Galerkin method \cite{meirovitch1994new} is reliable for numerically  solving such problems. 
It is now described.


Assume the expansion
\begin{equation}
\bar{y}_N (\bar{x},t) =  \mathbf{s}^\text{T}\mathbf{b},
\label{eq:appsol}
\end{equation}
where $\mathbf{s}(\bar{x}) = [\sin\left(n \pi \bar{x} \right)]^\text{T}$, for $n = 1\ldots N$, is an $N$-dimensional column vector of Fourier sine modes, and $\mathbf{{b}}(\bar{t})$ = $[b_1(\bar{t}),\,b_2(\bar{t}),\, \cdots\, b_N(\bar{t})]^\text{T}$ is a column vector of time-varying unknown coefficients. The modes individually vanish at the boundaries, thereby satisfying the geometric boundary conditions. Then (\ref{eq:nondim}) evaluated at $\bar{y} = \bar{y}_N$ results in an error, called the \emph{residue}. By setting the projections of the residue on the $N$ Fourier sine modes to zero yields a set of $N$ differential algebraic equation, which may be cast in the following state-space form:
\begin{subequations}
  \label{eq:mateqn}
\begin{equation}
\frac{d \mathbf{q} }{d \bar{t} } = \mathbf{A}  \mathbf{q},
\end{equation}
where the state vector and the state matrix are, respectively,  
\begin{eqnarray}
  \mathbf{q} = 
  \begin{bmatrix}
  \dot{\mathbf{b}}  \\ {\mathbf{b}}
  \end{bmatrix}  
  \text{\; and \;} 
\mathbf{A} = 
\int_0^1 \begin{bmatrix} -\mathbf{M}^{-1}\mathbf{C} & \mathbf{1} \\ -\mathbf{M}^{-1}\mathbf{K}   & \mathbf{0} \end{bmatrix} d\bar{x},
\end{eqnarray}
\end{subequations}
with $\mathbf{1}$ representing an \textit{N}$\times$\textit{N} identity matrix and $\mathbf{0}$ being the  \textit{N}$\times$\textit{N} zero matrix, while 
\refstepcounter{equation}
\label{eq:beamentries}
$$
\mathbf{M} = \left( \mathbf{s} \mathbf{s}^\text{T} \right) - \lambda^{-2}   \left( \mathbf{s'} \mathbf{s'}^\text{T} \right),  
\;
\mathbf{C} = c \left\lbrace \mathbf{s} \mathbf{s}^\text{T} \right\rbrace + 2 \bar{v} \left( \mathbf{s} \mathbf{s'}^\text{T} + \lambda \mathbf{s'} \mathbf{s''}^\text{T} \right) 
\eqno{(\theequation{ \text{a},\text{b} })}
$$
and
$$
\mathbf{K} = \left( 1 - \lambda^{-2} \bar{v}^2 \right) \,\left( \mathbf{s''} \mathbf{s''}^\text{T} \right) + \left( \bar{v}^2 - \bar{T}(\bar{x}) \right)\, \left( \mathbf{s'} \mathbf{s'}^\text{T} \right) + \frac{c \bar{v}}{2}  \left( \mathbf{s} \mathbf{s'}^\text{T} - \mathbf{s'} \mathbf{s}^\text{T} \right) 
\eqno{(\theequation{ \text{c} })}
$$
are the mass, damping and stiffness matrices, respectively, with the $(\,)\mathbf{'}$ denoting $d/d\bar{x}$. 
Note that the mass matrix $\mathbf{M}$ is symmetric, 
while the damping matrix $\mathbf{C}$ has a symmetric part $c \left( \mathbf{s} \mathbf{s}^\text{T} \right)$ and a skew-symmetric part $2 \bar{v} \left( \mathbf{s} \mathbf{s'}^\text{T} + \lambda \mathbf{s'} \mathbf{s''}^\text{T} \right)$ called the gyroscopic damping.  From the stiffness matrix  $\mathbf{K}$, after replacing for $\bar{T}$ from (\ref{eq:Tnondim}), we obtain four terms: the  material stiffness $\left(1 - \lambda^{-2} \bar{v}^2 \right) \,\left( \mathbf{s''} \mathbf{s''}^\text{T} \right)$, the geometric stiffness $\left( \bar{v}^2 - \mu^2 \right)\, \left( \mathbf{s'} \mathbf{s'}^\text{T} \right)$, the stiffness $\varrho \cos\varphi\, (1-\bar{x})\,\left( \mathbf{s'} \mathbf{s'}^\text{T} \right)$ due to the cable's weight,   
and the stiffness $c \bar{v}\, \left( \mathbf{s} \mathbf{s'}^\text{T} - \mathbf{s'} \mathbf{s}^\text{T} \right)/2$ due to damping. While the first three stiffness terms are symmetric, the last one is skew-symmetric. 
The state-space form for the string model (\ref{eq:nondimstring}) may be obtained by setting $\lambda^{-2}=0$ and $\mu=1$ in (\ref{eq:beamentries}). 

Equation (\ref{eq:mateqn}) may be treated as an \textit{eigenvalue} problem  for \textit{modal analysis}, or an initial value problem for  \emph{transient analysis}. In {modal analysis}, the cable's natural frequencies corresponding to the specified modes are extracted directly from (\ref{eq:mateqn}). They provide a qualitative insight into the system \cite{ziegler2013principles}. In a  transient analysis the solution $\mathbf{b}(\bar{t})$ is calculated through direct time-integration of the state-space equation (\ref{eq:mateqn}a) with given initial conditions $\mathbf{b}(0)$ and $\dot{\mathbf{b}}(0)$. The material displacements are then obtained from (\ref{eq:appsol}). 

We now discuss modal analysis in detail, followed by a brief discussion of transient analysis. 
\subsection{Modal Analysis}
\label{sec:modal}
An eigenvalue problem is formulated from (\ref{eq:nondim}) by setting ${\mathbf{b}}(\bar{t}) = [\exp(\omega_n \bar{t})]^\text{T}$,  
where $\omega_n$ is the eigenvalue (frequency) associated with the $n^\text{th}$ mode, assumed to be given by $\sin\left(n \pi \bar{x}\right)$. 
The state matrix $\mathbf{A}$ in (\ref{eq:mateqn}b) is \textit{not} symmetric in general, but it has complex conjugate eigenvalues. 
Accordingly, the transverse displacement is 
\begin{equation}
\bar{y}_N(\bar{x},\bar{t}) = \text{Re} \sum_{n=1}^{N}  \exp [ \{ \text{Re}(\omega_n) \pm i\, \text{Im}(\omega_n) \}  \bar{t} \, ] \sin({n \pi \bar{x}}),
\end{equation}
in which 
the $N$ sinusoidal modes collectively define the instantaneous shape of the cable, while the oscillatory part $\exp\{ \pm i\, \text{Im}(\omega_n) \bar{t} \}$ 
modulates the material point's vibration about its equilibrium state. The growth or decay of the amplitude of these vibrations is governed by $\exp\{ \text{Re}(\omega_n) \bar{t} \}$. 
The real part of the eigenvalue, thus, governs the solution's, and hence the cable's instability. The  eigenvalues depend upon the geometric and material parameters of the cable, 
\emph{viz}. travel speed and inclination of the cable,  tension and bending rigidity of the cable, and the cable's slenderness ratio. 

The accuracy of modal analysis depends on the number of modes we consider for approximating the solution, and  \ref{appendix-convergence} discusses this. 
\subsection{Stability}
Instability in operation corresponds to positive real part of the smallest eigenvalue.
\begin{figure}[h!]
\begin{center}
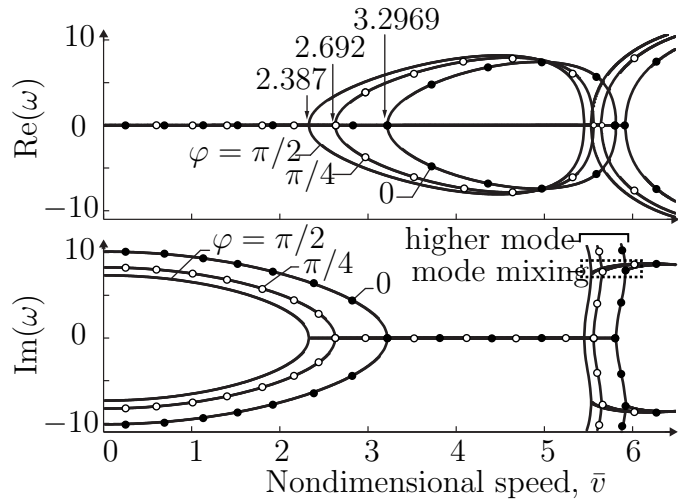
\end{center}
\caption{Comparison of the first eigenvalues for a cable modeled as a traveling  Euler-Bernoulli beam, inclined at various angles to gravity: $\varphi=\pi/2$ represents the vertically traveling cable and $\varphi=0$ is the horizontally traveling cable. The higher modes appear at later speeds and are observed to mix with the primary modes.}
\label{fig:freqspec}
\end{figure}
When this happens the displacements grow exponentially with time. This may result in failure and/or bring in nonlinear effects, hitherto ignored. The instability criterion identifies the \textit{critical value} of the governing parameter at which the real part of the smallest (first) eigenvalue first becomes positive. For example, in Fig.\,\ref{fig:freqspec}, from the bifurcation in $\text{Re}(\omega)$ we identify nondimensional speeds $\bar{v} = 2.387$, $2.692$ and $3.2969$ as the \textit{critical} speeds for cables traveling at inclinations $\varphi=\pi/2,\pi/4$ and $0$, respectively. Beyond these critical speeds $\text{Re}(\omega)$ splits into a positive and a negative part, whereas $\text{Im}(\omega)$ vanishes. 
With increase in inclination 
the bifurcation point shifts to the left. This implies a lowering of the critical speed of operation with increasing inclination leading to an earlier onset of instability. The critical speed $\bar{v}=3.2969$ for the horizontally traveling beam ($\varphi=0$) agrees well with the value reported in \cite{wickert1990classical}. 

\begin{figure}[!h]
  \begin{center}
   \tikzset{every mark/.append style={scale=1.25}}
   \pgfplotsset{
  every axis plot post/.append style={
  thick,
   every mark/.append style={fill=white}
   }}
  \begin{tikzpicture}
  \begin{axis}
   [
  cycle list name=mark list,
   xlabel={Nondimensional end tension, $\mu$},
    ylabel={Nondimensional speed, $\bar{v}$} , 
    ylabel style={sloped like y axis},
    xmin = -0.2, xmax = 5.2,
    ymin = 1.5, 
    ymax = 6.2,
    axis y line=left,
    axis x line=bottom,
   table/x = phi_pi16,
  y tick label style={
        /pgf/number format/.cd,
            fixed,
            fixed zerofill,
            axis y discontinuity=crunch,
            precision=0,
        /tikz/.cd
    },
     legend style={at={(1,0.4)},draw=none},
     title={\underline{Inclination, $\varphi$}},
     title style={xshift=2.6cm,yshift=-3.6cm},
  ]
\addplot table[ y=8,col sep=comma] {\VcphimuT} ; \addlegendentry{$0$}
\addplot table[ y=6,col sep=comma] {\VcphimuT} ; \addlegendentry{$\pi/8$}
\addplot table[ y=4,col sep=comma] {\VcphimuT} ; \addlegendentry{$\pi/4$}
\addplot table[ y=0,col sep=comma] {\VcphimuT}; \addlegendentry{$\pi/2$}
%
\node at (axis cs:2,2.0) [black,fill=none,below=0,rotate=0] {Stable};
\node at (axis cs:2,5.8) [black,fill=none,below=0,rotate=0] {Unstable};
\node at (axis cs:2,2.9) [black,fill=none,below=0,rotate=35] {(Vertical)};
\node at (axis cs:2,3.8) [black,fill=none,above=0,rotate=30] {(Horizontal)};
 \end{axis}
 \end{tikzpicture}
  \end{center}
  \caption{Stability curves for traveling Euler-Bernoulli beams at various inclinations $\varphi$ are obtained by plotting the nondimensional speed $\bar{v}$ as a function of nondimensional end tension $\mu$. The parameter $\varrho = 10$.} 
  \label{fig:vcmu}
 \end{figure}
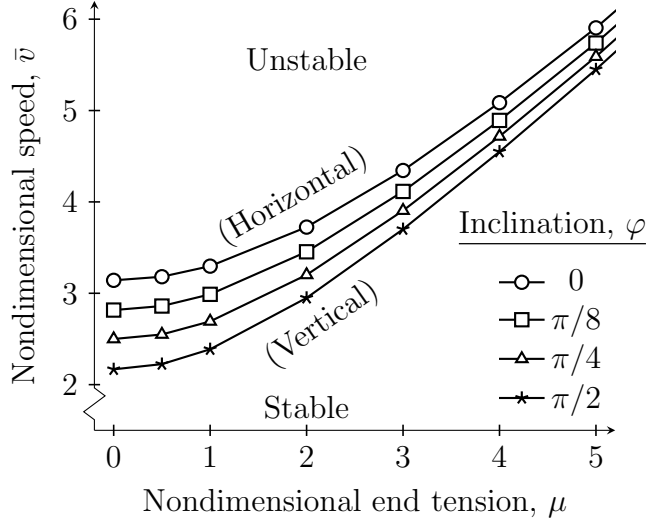

A higher tension in the cable stabilizes the system. This role of tension in enhancing stability is lowered by the action of gravity in inclined cables. This is illustrated in Fig.\,\ref{fig:vcmu} that plots critical travel speed as a function of the dimensionless end tension $\mu$ for various inclinations $\varphi$. 
The resulting curve for a given $\varphi$ is the boundary separating regions of stable and unstable operations, which lie below and above the curve, respectively. Stability curves at intermediate values of inclination  lie between those belonging to vertically and horizontally traveling cables.
 Furthermore, we observe that, as expected, increasing the cable tension delays the onset of instability,  thereby ensuring stable operation at relatively higher speeds. 
 Finally, we see that a more inclined cable becomes unstable at lower speeds. 
  The preceeding discussion is illustrated by the following example.
\paragraph{Example 1}\label{para:ex1}
A 0.5 mm diameter steel cable ($E=210$ GPa, $\rho=7800$ Kg/m$^3$), traveling horizontally ($\varphi=0$) between rollers which are $L=$ 1 m apart, is stable when the end tension $T(L)$ is 1.75 mN ($\mu = 1$) and the speed is about 3.26 m/s  ($\bar{v} = 3.29$); cf. Fig.\,\ref{fig:vcmu}. However, an inclined  cable ($\varphi>0$) under the same $T(L)$ is unstable at that speed. It becomes stable at lesser speeds: 2.95 m/s ($\bar{v} = 2.98$)  when $\varphi=\pi/8$, 2.66  m/s ($\bar{v} = 2.69$) when $\varphi=\pi/4$ and 2.35 m/s ($\bar{v} = 2.38$) when traveling vertically ($\varphi=\pi/2$).  Thus, inclination has a destabilizing effect. 

\begin{figure}[h!]
  \begin{center}
   \tikzset{every mark/.append style={scale=1.25}}
   \pgfplotsset{
  every axis plot post/.append style={
   every mark/.append style={fill=white}
   }}
  \begin{tikzpicture}
  \begin{axis}
   [smooth,
  cycle list name=linestyles,
   xlabel= {Slenderness ratio, $\lambda$},
   ylabel={Nondimensional speed, $\bar{v}$} , 
  axis x line = bottom,
  axis y line = left, 
  xmin = 5, xmax = 20,
  ymin = 2.5,ymax = 3.5,
  table/x expr = \thisrow{phi_pi16}^-0.5,
  y tick label style={
        /pgf/number format/.cd,
            fixed,
            fixed zerofill,
            precision=1,
        /tikz/.cd
    },
  ]
 \addplot [thick,mark=*] table[y=0,col sep=comma] {\VcphilamT} 
 %
(axis cs:16,3) node [white,fill=none,rotate=0] {(Rayleigh)};
 \addplot [very thick, dashed] table[ y=0,col sep=comma] {\VcphilamEBT} 
 %
(axis cs:16,3.2) node [white,fill=none,rotate=0] {(Euler-Bernoulli)};
\addplot+[nodes near coords,enlargelimits=0.2,only marks, mark = none, point meta=explicit symbolic] coordinates 
   {
        (14,2.7) [Stable]
       (14,3.3) [Unstable]
    };
 \end{axis}
 \end{tikzpicture}
 \end{center}
 \caption{ Stability curves obtained by plotting nondimensional speed $\bar{v}$ as a function of slenderness ratio $\lambda$ for horizontally traveling ($\varphi=0$) Rayleigh (solid line) and Euler-Bernoulli (dashed line) beams.}
 \label{fig:vclam} 
 \end{figure}
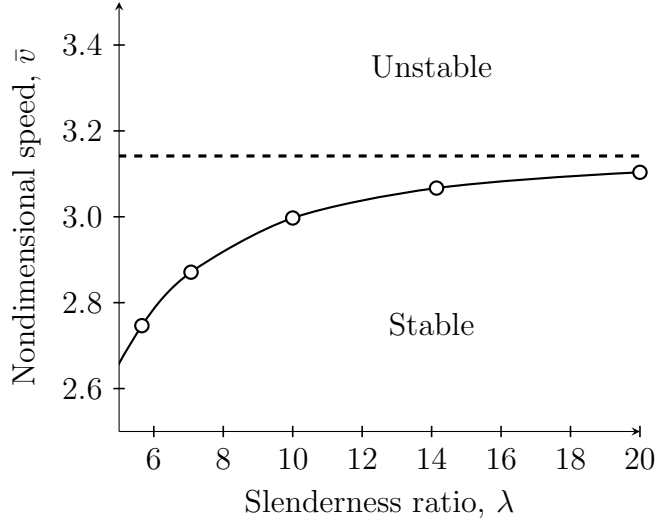

Increasing the tension to $T(1\text{m}) = 7$ mN ($\mu  = 2$) increases the critical speeds considerably: 3.68 m/s at $\varphi=0$, 3.42 m/s at $\varphi=\pi/8$, 3.17  m/s at  $\varphi=\pi/4$ and 2.92 m/s at $\varphi=\pi/2$.  However, with further increase in tension to $T(1\text{m}) = 28$ mN ($\mu  = 4$) the respective critical speeds: 5.04 m/s at $\varphi=0$, 4.84 m/s at $\varphi=\pi/8$, 4.67  m/s at  $\varphi=\pi/4$ and 4.50 m/s at $\varphi=\pi/2$, are closer to each other. The critical speeds for all $\varphi$ converge while rising further; see Fig.\,\ref{fig:vcmu}. Thus, tension contributes to the stability.
%
%
\subparagraph*{ }
Consider now a cable modeled as a Rayleigh beam wherein rotary inertia is retained. In a horizontally traveling Rayleigh beam the slenderness ratio $\lambda$ improves the stability for any nonzero end-tension $\mu$ and mass density $\varrho$, as shown by the solid curve in Fig.\,\ref{fig:vclam}. At higher $\lambda$ the rotary inertia term in (\ref{eq:nondim}) becomes small compared to the bending rigidity. Therefore, with increasing slenderness ratio, the stability boundary of the Rayleigh beam approaches that of a traveling Euler-Bernoulli beam (dashed line); the latter, of course,  remains unaffected by the beam's slenderness when $\varphi=0$.
%

\begin{figure}[h!]
  \begin{center}
   \tikzset{every mark/.append style={scale=1.25}}
   \pgfplotsset{
  every axis plot post/.append style={
  thick,
   every mark/.append style={fill=white}
   }}
  \begin{tikzpicture}
  \begin{axis}[
  cycle list name=mark list,
  xlabel={Inclination, $\varphi$ }, 
   ylabel={Nondimensional speed, $\bar{v}$} , 
   width=8.3cm,
   height=4.0cm,
   axis x line = bottom,
   axis y line = left,
    y tick label style={
        /pgf/number format/.cd,
            fixed,
            fixed zerofill,
            precision=5,
        /tikz/.cd
    },
    axis y discontinuity=crunch,
     table/x = phi_pi16, 
     xmin = -0.2, xmax=9,
     ymin = 3.141585, ymax = 3.1416105,
     xticklabels={o,$0$,$\pi/8$,$\pi/4$,$3\pi/8$,$\pi/2$},
     ytick={3.14159,3.1416,3.14161},
     legend style={legend pos=outer north east,draw=none},
     title={\underline{Slenderness ratio, $\lambda$}},
     title style={xshift=3.7cm,yshift=-0.26cm},
    ]
  %
 \addplot table[ y=0.00125,col sep=comma] {\VcphilamEB}; \addlegendentry{25}
\addplot table[ y=0.0025,col sep=comma] {\VcphilamEB}; \addlegendentry{20}
\addplot table[ y=0.01,col sep=comma] {\VcphilamEB}; \addlegendentry{10}
\addplot table[ y=0.5,col sep=comma]  {\VcphilamEB}; \addlegendentry{1.5}
%
%
\addplot+[nodes near coords,enlargelimits=0.2,only marks, mark = none, point meta=explicit symbolic] coordinates 
   {
        (5,3.141586) [Stable]
       (5,3.141607) [Unstable]
    };
 \end{axis}
 \end{tikzpicture}
 \end{center}
  \caption{Variation with inclination $\varphi$ of critical nondimensional speed $\bar{v}_\text{crit}$ for cables modeled as Euler-Bernoulli beams. Several slenderness ratios $\lambda$ are considered. Slender cables are slightly more stable for all non-zero inclinations.}
\label{fig:vcphiEB}
\end{figure}
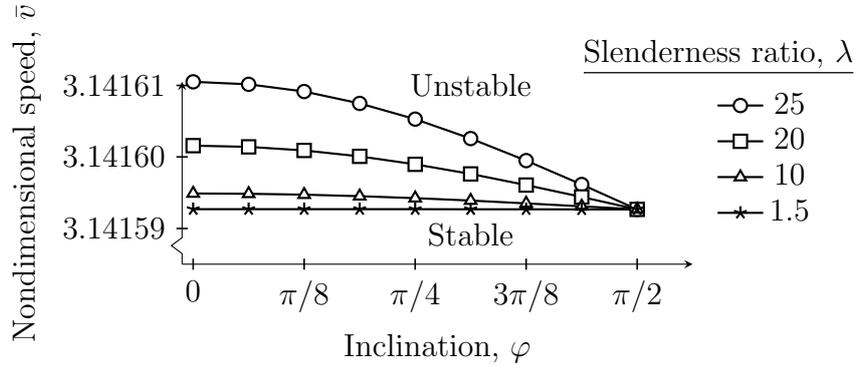

In passing, we note from Fig.\,\ref{fig:vcphiEB} that more inclined ($\varphi>0$) traveling Euler-Bernoulli beams with higher $\lambda$ are slightly less stable than the horizontal ones. 
 However, when $\lambda$ is small, the inclination barely affects the Euler-Bernoulli beam's stability. 
%
%

\begin{figure}[h!]
\begin{center}
 %
%
%
\def\svgwidth{8.5cm}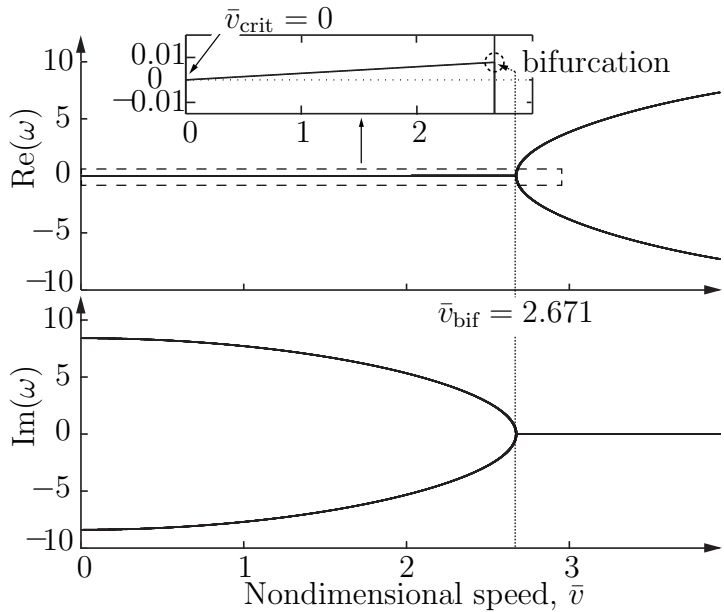
\end{center}
\caption{Real and imaginary part of the smallest (first) eigenvalue of an undamped ($c=0$),  inclined $(\varphi=\pi/4)$, traveling Rayleigh beam are  plotted as functions of speed. Other parameters are $\varrho=10$, $\mu = 1$ and $\lambda=5$. The inset shows the existence of positive real part at all nondimensional speeds $\bar{v}>0$, while the bifurcation (as in Fig.\,\ref{fig:freqspec}) happens much later when $\bar{v} = \bar{v}_\text{bif}=2.671$. } 
\label{fig:rayleighspec}
\end{figure}

Surprisingly, we find that an inclined ($\varphi > 0$) traveling Rayleigh beam is inherently unstable, i.e.  at any $\varphi > 0$  
the real part of the smallest eigenvalue is always positive, as shown in the inset of Fig.\,\ref{fig:rayleighspec} for the case when $\varphi=\pi/4$. Moreover, unlike an Euler-Bernoulli beam (Fig.\,\ref{fig:freqspec}), the critical speed is different from the bifurcation speed $\bar{v}_\text{bif} = 2.671$ at which the smallest Re($\omega$) bifurcates and the associated Im($\omega$) vanishes.   

\begin{figure}[h!]
\begin{center}
\def\svgwidth{8.5cm}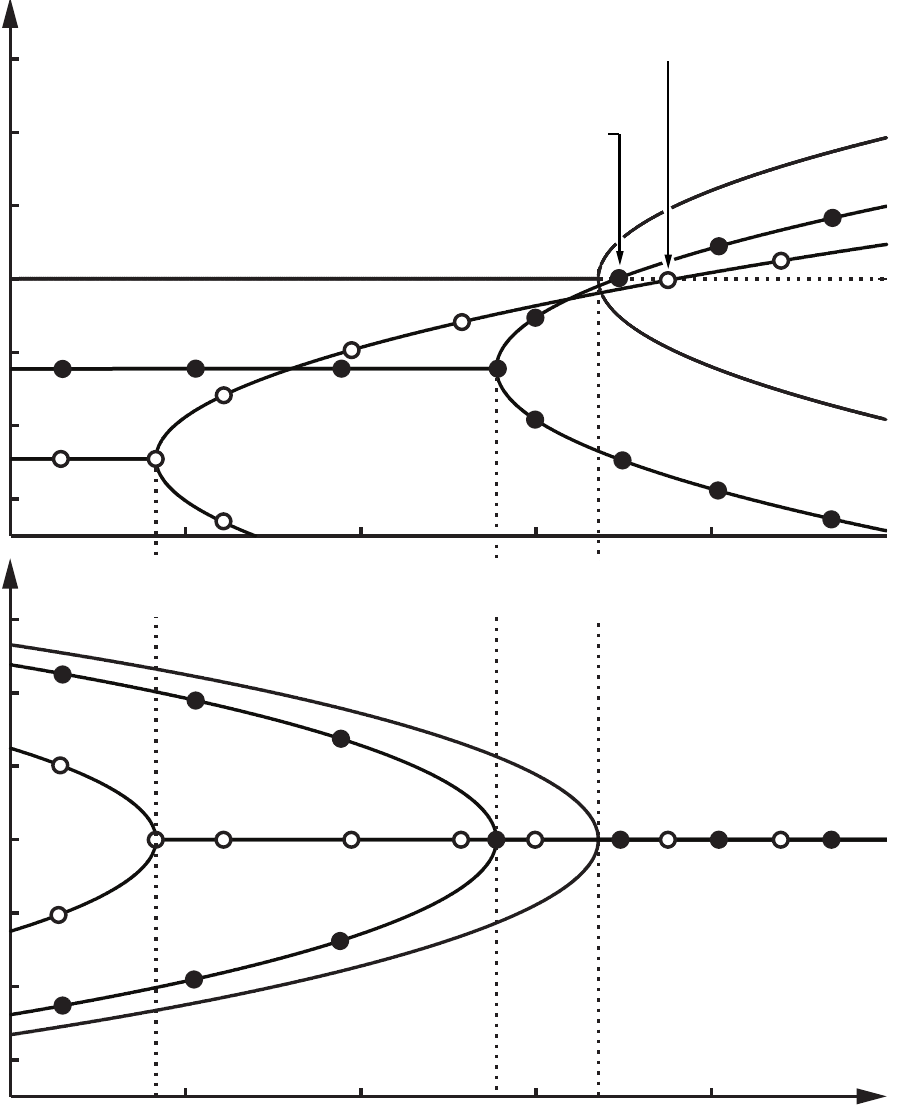
\end{center}
\caption{Real and imaginary parts of the first mode's eigenvalue $\omega$ for a  damped, traveling Rayleigh beam are shown as functions of nondimensional speed $\bar{v}$ for these values of nondimensional viscosity: $c=0$, $5$ and $10$. We observe that with damping the critical speed $\bar{v}_\text{crit}$ increases from 0 for $c=0$ to $2.692$ for $c=5$,  and $2.759$ for $c=10$. Also,   $\text{Re}(\omega)$ bifurcates at speeds $\bar{v}_\text{bif}<\bar{v}_\text{crit}$ for $c=5$ and $10$, in contrast to when $c=0$. }
\label{fig:rayleighspec0510}
\end{figure}
However, we observe that the inclusion of damping ($c>0$) 
shifts the real part of the first eigenvalue below zero, thereby removing the inherent instability of a traveling Rayleigh beam. The amount that Re($\omega$) shifts depends upon the amount of damping $c$ introduced. For example, when $c=5$ (Fig.\,\ref{fig:rayleighspec0510}) we identify $\bar{v}_\text{crit} = 2.692$ as the critical speed at which a real part of the smallest eigenvalues first begins to appear positive. More damping shifts $\bar{v}_\text{crit}$ further to the right. This shift in critical speed due to damping  maybe employed to provide a suitable range of speeds $\bar{v}<\bar{v}_\text{crit}$ for stable operation. However, in the damped traveling Rayleigh beams of Fig.\,\ref{fig:rayleighspec0510} 
the speed $\bar{v}_\text{bif}$ at which the first eigenvalue  bifurcates 
is less than $\bar{v}_\text{crit}$. This indicates the existence of complex eigenvalues at sub-bifurcation  speeds, resulting in underdamped oscillations of the beam in its  first mode during stable operation. We note that, in contrast to $\bar{v}_\text{crit}$, the bifurcation point in the damped beam shifts left, i.e. $\bar{v}_\text{bif}$ decreases with increase in damping. 
We note from this behaviour of  $\bar{v}_\text{bif}$ that when $c$ is increased enough, at a given  $\bar{v}$  eventually $\text{Im}(\omega)$ of the first mode  will cease to appear. When this happens, the higher modes (not shown), similar to those found in Euler-Bernoulli beams (Fig.\,\ref{fig:freqspec}), become significant and govern the beam's oscillations.

Finally, we note that 
a damped Euler-Bernoulli beam shows a similar behaviour. 
\subsubsection*{String model}
\begin{figure}[h!]
  \begin{center}
   \tikzset{every mark/.append style={scale=1.25}}
   \pgfplotsset{
  every axis plot post/.append style={
  thick,
   every mark/.append style={fill=white}
   }}
  \begin{tikzpicture}
  \begin{axis}[
  cycle list name=mark list,
  name=plot1,width=7.5cm,height=7.0cm,
    domain=0:1,
  xlabel={Inclination, $\varphi$ }, 
   ylabel={Nondimensional speed, $\bar{v}$} , 
   axis x line = bottom,
   axis y line = left,
    y tick label style={
        /pgf/number format/.cd,
            fixed,
            fixed zerofill,
            precision=1,
        /tikz/.cd
    },
     table/x expr = {8-\thisrow{phi_pi16}},
     xmin = 0, xmax=9,
     ymin= 0, ymax = 1.2,
     xticklabels={o,$0$,$\pi/8$,$\pi/4$,$3\pi/8$,$\pi/2$},
     legend style={at={(0.33,0.44)},draw=none},
     title={(\textbf{a})},
     title style={xshift=1.8cm,yshift=-0.7cm},
    ]
\node at (axis cs: 2.9,0.72) [anchor=north east,fill=none] { {Scaled end}};
  \node at (axis cs: 2.9,0.64) [anchor=north east,fill=none] { \underline{tension $\mu_\text{s}$} };
 \addplot table[ y=1,col sep=comma] {\Vcphirho}; \addlegendentry{$1$}%
\addplot table[ y=2,col sep=comma] {\Vcphirho}; \addlegendentry{$2$} %
\addplot table[ y=4,col sep=comma] {\Vcphirho}; \addlegendentry{$4$} %
\addplot table[ y=10,col sep=comma]  {\Vcphirho}; \addlegendentry{$10$} %
\addplot table[ y=1000,col sep=comma]  {\Vcphirho}; \addlegendentry{$1000$} %
\node at (axis cs:5,0.4) [anchor=north east,fill=none] {Stable};
\node at (axis cs:5,1.2) [anchor=north east,fill=none] {Unstable};
%
 \end{axis}
 \begin{axis}[
  cycle list name=mark list,
  name=plot2,at={($(plot1.east)+(0.5cm,0)$)},anchor=west,width=7.5cm,height=7.0cm,
    anchor=origin, 
    xshift=1.5cm,
    yshift= -2.7cm,
    domain=0:1,
  xlabel={Scaled end tension, $\mu_\text{s}$ }, 
   ylabel={Nondimensional speed, $\bar{v}$} , 
   axis x line = bottom,
   axis y line = left,
    y tick label style={
        /pgf/number format/.cd,
            fixed,
            fixed zerofill,
            precision=1,
        /tikz/.cd
    },
     table/x = phi_pi16,
     xmin = 0, xmax = 12,
     ymin= 0.4, 
     ymax = 1.1,
     legend style={legend pos=south east,draw=none},
     title={(\textbf{b})},
    title style={xshift=2cm,yshift=-0.7cm},
    ]
  \node at (axis cs: 12,0.76) [anchor=north east,fill=none] {\underline{Inclination, ${\varphi}$ }};
  %
 %
 \addplot table[ y expr=\thisrow{8}^0.5,col sep=comma] {\VcphirhoT} ; \addlegendentry{$0$}%
\addplot table[ y expr=\thisrow{6}^0.5,col sep=comma] {\VcphirhoT}; \addlegendentry{$\pi/8$} %
\addplot table[ y expr=\thisrow{4}^0.5,col sep=comma] {\VcphirhoT}; \addlegendentry{$\pi/4$} %
\addplot table[ y expr=\thisrow{0}^0.5,col sep=comma]  {\VcphirhoT}; \addlegendentry{$\pi/2$} %
\node at (axis cs:6,0.6) [anchor=north east,fill=none] {Stable};
\node at (axis cs:6,1.1) [anchor=north east,fill=none] {Unstable};
%
 \end{axis}
 \end{tikzpicture}
 \end{center}
  \caption{ (\textbf{a}): Nondimensional speed $\bar{v}$ of traveling strings as a function of their inclination $\varphi$ for several choices of scaled end tension $\mu_\text{s}$.  (\textbf{b}): Nondimensional speed $\bar{v}$ of traveling strings as a function of their scaled end tension $\mu_\text{s}$  for several values of inclination $\varphi$.}
\label{fig:vcphirho}
\end{figure}
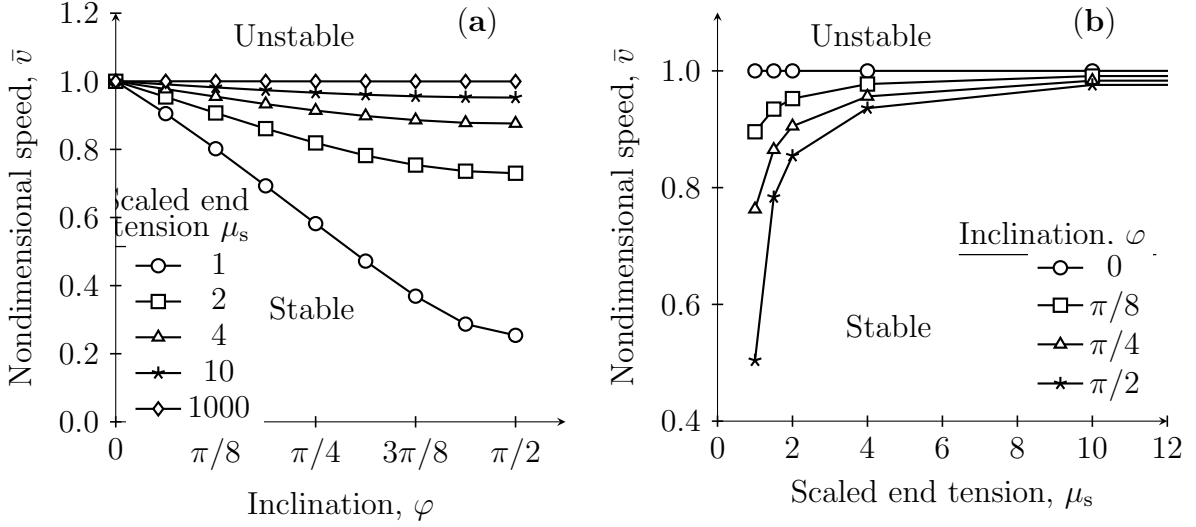

For highly flexible cables (negligible bending rigidity), the tension in the cable and its inclination govern stability. 
Figure \ref{fig:vcphirho}(a) plots travel speed $\bar{v}$ as a function of inclination $\varphi$. The resulting stability curves correspond to different values of the scaled end tension 
\[\mu_\text{s} = \dfrac{ T(L)|_{\varphi} }{ T(L)|_{0} } ,\] 
which is a ratio of end tension $T(L)$ at inclination $\varphi$ to that at $\varphi=0$ (horizontal). We again note that less inclined cables are more unstable, and tension enhances stability. 
However, unlike for beams, the tension does \textit{not} affect the critical speed of horizontally traveling strings. In this case the critical speed is 1,  matching the analytic solution in  \cite{hagedorn2007vibrations}. Figure \ref{fig:vcphirho}(b), on the other hand, plots $\bar{v}$ as a function of the scaled end tension and each curve corresponds to a different $\varphi$, similar to Fig.\,\ref{fig:vcmu} for traveling Euler-Bernoulli beams. Although not as transparent as Fig.\,\ref{fig:vcphirho}(a), we draw similar  conclusions from Fig.\,\ref{fig:vcphirho}(b). However, it is important to note that in contrast to beams (Fig.\,\ref{fig:vcmu}), the stability curves of strings in Fig.\,\ref{fig:vcphirho}(b)  are concave and they all converge to the critical speed of a horizontal string at high end-tensions. 

The following example illustrates this discussion. 
%
%
%
\paragraph{Example 2}\label{para:ex2}
The 0.5 mm steel cable mentioned in \nameref{para:ex1}, when modeled as a traveling string, is stable for $T(L=1\text{m}) = 1.75$ mN ($\mu_\text{s}  = 1$) at travel speeds less than 1 m/s  ($\bar{v} = 1$) when traveling horizontally ($\varphi=0$), 0.8 m/s ($\bar{v} = 0.802$)  when $\varphi=\pi/8$, 0.57 m/s ($\bar{v} = 0.582$) when $\varphi=\pi/4$ and 0.25 m/s ($\bar{v} = 0.254$) when traveling vertically ($\varphi=\pi/2$).

Thus, under the same end conditions, a traveling cable when modeled as string predicts a lesser critical speed than when it is modeled as an Euler-Bernoulli beam.  This shows the stabilizing effect of bending rigidity $EI$ in the beam. We further note that the difference between the critical speeds of Euler-Bernoulli beam and  the string increases with the inclination $\varphi$. For this particular  steel cable, the $v_\text{crit}$ of the horizontally traveling Euler-Bernoulli beam (\nameref{para:ex1}) is more than three times that of the horizontally traveling string (\nameref{para:ex2}), while its almost ten when the cable is traveling vertically. Finally, higher end tension, for example $T(1\text{m}) = 7$ mN ($\mu  = 2, \mu_\text{s}=1$) or $T(1\text{m}) = 28$ mN ($\mu  = 4, \mu_\text{s}=1$), considerably reduces the difference in their respective critical speeds. However, any further increase in $T(L)$ doesn't reduce the difference much, and the $v_\text{crit}^\text{EB beam} > v_\text{crit}^\text{String}$ at all $\varphi$.

\section{Energetics} \label{sec-energetics}
\pgfplotstableread[col sep = comma]{timeseriesH.txt}\timeH
\pgfplotstableread[col sep = comma]{timeseriesV.txt}\timeV
Further insight into the system is provided by investigating the manner in which energy flows into and away from a traveling cable. This also has practical design implications.  
In this section we examine the flow of energy in traveling Euler-Bernoulli beams below and above the critical speed. 

The total mechanical energy per unit length of the beam is 
\[
\bar{E}(\bar{x},\bar{t}) = \bar{E}_K(\bar{x},\bar{t}) + \bar{E}_P(\bar{x},\bar{t}),
\]
{where the kinetic energy density} 
\[
\bar{E}_K(\bar{x},\bar{t}) = \frac{1}{2}(\bar{v}^2 + \dot{\bar{y}}^2) 
\]
{and the potential energy density} 
\[
\bar{E}_P(\bar{x},\bar{t}) = \frac{1}{2}(\bar{\partial_x}^2 \bar{y})^2 + \frac{\bar{T}(\bar{x})}{2} (\bar{\partial_x} \bar{y})^2 + \varrho \bar{x} \sin\varphi. 
\]
The total rate of change of mechanical energy is 
%
\begin{align}
\dot{\bar{E}} (\bar{t}) &= \int\limits_0^1 \bar{\partial_t} {\bar{E}} + \bar{v}\,\bar{\partial_x} \bar{E} \, d\bar{x}
=\left| \left\lbrace \bar{T}(\bar{x}) \,\bar{\partial}_x \bar{y} - \bar{\partial}^3_x \bar{y} \right\rbrace \left( \bar{v} \,\bar{\partial}_x \bar{y} \right) \right|^1_0 +  \int\limits_0^1  \left( \varrho \sin\varphi\, \bar{\partial}_x \bar{y} \right) \left(\bar{v} \, \bar{\partial}_x \bar{y} \right) d\bar{x},
 \label{eq:EdotV}
\end{align}
where the first term is the power supplied to the cable across the end supports by the action of the  transverse component of tension and shear force. This term estimates the inflow and outflow of energy from the boundaries due to the presence of non-zero convective velocity $\bar{v} \,\bar{\partial}_x \bar{y}$ there; note that the local velocity $\bar{\partial}_t \bar{y}$ vanishes at the supports. The second term in \eqref{eq:EdotV} is the contribution of the cable's self-weight, due to its inclination, which vanishes for a horizontally traveling cable.

The time rate of change of energy (\ref{eq:EdotV}) is zero in a non-translating ($\bar{v}=0$) cable, 
 but not if the cable is translating. 
In this the traveling cable exhibits a \textit{non-conservative} behaviour. This behaviour, however, doesn't imply that the traveling cable is always unstable. 
Recall that through modal analysis (Sec.\,\ref{sec:modal}) it was possible to identify a range of travel speeds and other system parameters for stable operations. 
Therefore, in the following, we study the stability of horizontally and vertically traveling cable by monitoring the evolution of the total mechanical energy and the transverse displacement of a material particle passing through the  mid-span ($x=L/2$) of the cable.  
%
\subsection*{Transverse displacement and the rate of change of energy}
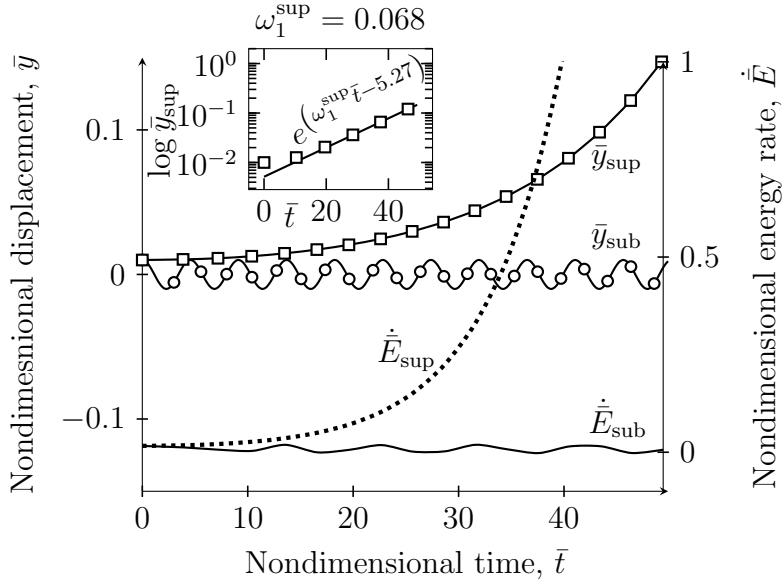
\begin{figure}[h!]
\begin{center}
 \tikzset{every mark/.append style={scale=1},solid}
   \pgfplotsset{
  every axis plot post/.append style={
   black,
   }}
\begin{tikzpicture}
\begin{axis}[smooth,
              ymin=-0.15,ymax=0.15,
               ytick={-0.1,0,0.1},
               xlabel={Nondimensional time, $\bar{t}$}, 
                ylabel={Nondimesnional displacement, $\bar{y}$},
                axis y line=left, 
               axis x line=bottom, 
              width=8.5cm,
               ]
 %
%
  \addplot [mark=*,mark options={fill=white, mark repeat=15},solid,thick] table[ x = T3p26, y = Y3p26, col sep=comma] {\timeH};
\addplot [mark=square*,mark options={fill=white},solid,thick] table[ x = t3p27, y = y3p27, col sep=comma] {\timeH};
%
%
%
\node at (axis cs:45,0.1) [black,fill=none,below=2,rotate=0] {$\bar{y}_\text{sup}$};
\node at (axis cs:45,0.01) [black,fill=none,above=0,rotate=0] {$\bar{y}_\text{sub}$};
\node at (axis cs:14,0.025) [black,fill=none,above=0,rotate=0] {$\bar{t}$}; 
\node at (axis cs:4.5,0.1) [black,fill=none,above=0,rotate=90] {$\log \bar{y}_\text{sup}$};
 \end{axis}
\begin{axis}[smooth,
              ymin=-0.1,ymax=1,
               ytick={0,0.5,1},
                ylabel={Nondimensional energy rate, $\dot{\bar{E}}$},
                 axis y line=right, 
               axis x line=bottom,
               hide x axis,
               ]
%
%
\addplot [mark=none,mark options={fill=white},solid,thick] table[ x = t3p26, y = er3p26, col sep=comma] {\timeH};
\addplot [mark=none,mark options={fill=white},dotted,ultra thick] table[ x = t3p27, y = er3p27, col sep=comma] {\timeH};
%
%
%
\node at (axis cs:25,0.18) [black,fill=none,above=0,rotate=0] {$\dot{\bar{E}}_\text{sup}$};
\node at (axis cs:45,0.01) [black,fill=none,above=0,rotate=0] {$\dot{\bar{E}}_\text{sub}$};
\end{axis}
\begin{semilogyaxis}[xshift=1.4cm,yshift=4cm,width=0.25\textwidth,title={$\omega_1^\text{sup} = 0.068$}, title style={xshift=0cm,yshift=-0.2cm,rotate=0}, ymax=2] 
%
 \addplot [mark=square*,mark options={fill=white,mark repeat=3}, only marks, thick] table[ x = t3p27, y = y3p27, col sep=comma] {\timeH} ; 
 \addplot [black,thick] table[ x = t3p27, y expr = exp(0.068*\thisrow{t3p27}-5.27), col sep=comma] {\timeH} node [pos=0.7, above=1, sloped] {$e^{\left( \omega_1^\text{sup} \bar{t} - 5.27 \right)} $ }; 
 \end{semilogyaxis}
\end{tikzpicture}
\end{center}
\caption{Temporal evolution of the transverse nondimensional displacement of the horizontally traveling cable at $\bar{x}=0.5$ ($x=L/2$) and the rate of change of total mechanical energy (nondimensional). The sub-critical and super-critical nondimensional speeds are $\bar{v}_\text{sub} = 3.26$ and $\bar{v}_\text{sup} = 3.27$, respectively. The inset semi-log plot shows that $\bar{y}_\text{sup}$ (~$\square$~) grows modestly for some period following which the  growth is exponential (---) at a rate $\omega_1^\text{sup} = \text{Re}\left( \omega_1^\text{sup} \right) = 0.068$.} 
\label{fig:displenergyH}
 \end{figure}

\begin{figure}[h!]
\begin{center}
 \tikzset{every mark/.append style={scale=1},solid}
   \pgfplotsset{
  every axis plot post/.append style={
   black,
   }}
\begin{tikzpicture}
\begin{axis}[smooth,
              ymin=-0.15,ymax=0.15,
               xlabel={Nondimensional time, $\bar{t}$}, 
                ylabel={Nondimensional displacement, $\bar{y}$},
                axis y line=left, 
               axis x line=bottom, 
              width=8.5cm,
               ]
 %
%
\node at (axis cs:10,0.095) [black,fill=none,below=0,rotate=0] {$\bar{y}_\text{sup}$};
\node at (axis cs:22,0.01) [black,fill=none,above=0,rotate=0] {$\bar{y}_\text{sub}$};
\node at (axis cs:40,0.02) [black,fill=none,above=0,rotate=0] {$\bar{t}$};
\node at (axis cs:24.5,0.1) [black,fill=none,above=0,rotate=90] {$\log \bar{y}_\text{sup}$};
%
%
%
\addplot [mark=square*,mark options={fill=white,mark repeat=4},solid,thick,smooth] table[ x = T2p38723, y = Y2p38723, col sep=comma] {\timeV}; 
%
 \addplot [mark=*,mark options={fill=white, mark repeat=15},solid,thick,smooth] table[ x = T2p37, y = Y2p37, col sep=comma] {\timeV};
\end{axis}
\begin{axis}[smooth,
              xmax=49,
              ymin=-0.1,ymax=1,
               ytick={0,0.5,1},
                ylabel={Nondimensional energy rate, $\dot{\bar{E}}$},
                 axis y line=right, 
               axis x line=bottom,
               hide x axis,
               ]
%
\node at (axis cs:6,0.2) [black,fill=none,above=0,rotate=0] {$\dot{\bar{E}}_\text{sup}$};
\node at (axis cs:14,0.0) [black,fill=none,above=0,rotate=0] {$\dot{\bar{E}}_\text{sub}$};
%
%
%
\addplot [mark=none,mark options={fill=white},solid,thick,smooth] table[ x = t2p37, y expr = +\thisrow{er2p37}, col sep=comma] {\timeV}; 
\addplot [mark=none,mark options={fill=white},dotted,ultra thick,smooth] table[ x = T2p38723, y expr = +\thisrow{ER2p38723}, col sep=comma] {\timeV}; 
%
\end{axis}
\begin{semilogyaxis}[xshift=4.15cm,yshift=4.0cm,width=0.25\textwidth,title={$\omega_1^\text{sup} = 0.068$},title style={xshift=0cm,yshift=-0.2cm,rotate=0},axis background/.style={fill=none}];
\addplot [mark=square*,mark options={fill=white,mark repeat=10},only marks,thick] table[ x = T2p38723, y = Y2p38723, col sep=comma] {\timeV};
 \addplot [black,thick] table[ x = T2p38723, y expr = exp(0.068*\thisrow{T2p38723}-3.3), col sep=comma] {\timeV} node [pos=0.4, above=0, sloped] {$e^{\left( \omega_1^\text{sup} \bar{t} - 3.3 \right)} $ };
\end{semilogyaxis}
 %
\end{tikzpicture}
\end{center}
\caption{Temporal evolution of the transverse nondimensional displacement of the vertically traveling cable at $\bar{x} = 0.5$ ($x=L/2$) and the rate of change of total mechanical energy (nondimensional). The sub-critical and super-critical nondimensional speeds are $\bar{v}_\text{sub} = 2.37$ and $\bar{v}_\text{sup} = 2.38723$, respectively. The inset semi-log plot shows that $\bar{y}_\text{sup}$ (~$\square$~) grows exponentially,  at a rate $\omega_1^\text{sup} = \text{Re}\left( \omega_1^\text{sup} \right) = 0.068$, after a rapid initial rise in its amplitude. }
\label{fig:displenergyV}
 \end{figure}
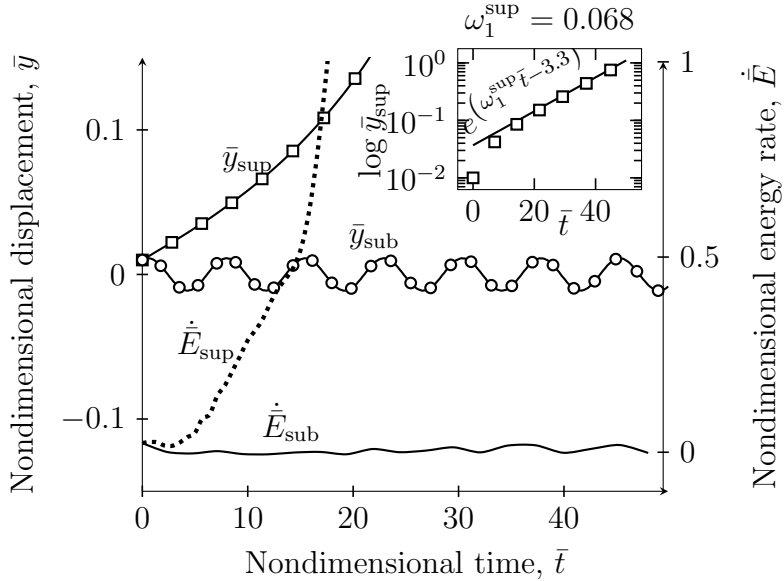

A standard explicit (Runge-Kutta) numerical time-integration of the state space equation (\ref{eq:mateqn}) is performed to obtain the temporal coefficients $\mathbf{b}(t)$. The traveling cable is assumed to be released from a displaced configuration corresponding to its first mode. Accordingly, the initial conditions are: $\mathbf{b}(0) = [0.01,\,0,\,0,\, \ldots\,]^\text{T}$ and $\mathbf{\dot{b}}(0) = [0,\,0,\,0,\, \ldots\,]^\text{T}$. 
The transverse displacement at $\bar{x}=0.5$ is then calculated from (\ref{eq:appsol}), and the rate of change of total energy $\dot{\bar{E}}$ is obtained from (\ref{eq:EdotV}). Figures\,\ref{fig:displenergyH} and \ref{fig:displenergyV} show the evolution of transverse displacement and $\dot{\bar{E}}$ for a horizontally and vertically ($\varphi=\pi/2$) traveling Euler-Bernoulli beam, respectively. 
The results at sub-critical and super-critical operations are compared keeping the critical speed corresponding to $\mu=1$ and $\varrho=10$ from Fig.\,\ref{fig:vcmu} as reference. The displacement ${\bar{y}}_\text{sub}$ and rate of change of energy $\dot{\bar{E}}_\text{sub}$ correspond to operation at sub-critical speed ${\bar{v}}_\text{sub} <  {\bar{v}}_\text{crit}$, 
while ${\bar{y}}_\text{sup}$ and $\dot{\bar{E}}_\text{sup}$ are calculated at super-critical speed ${\bar{v}}_\text{sup} > {\bar{v}}_\text{crit}$. 

The sub-critical $\dot{\bar{E}}_\text{sub}$ and ${\bar{y}}_\text{sub}$ are oscillatory and bounded over an extended period of time (see Fig.\,\ref{fig:displenergyH}). 
This is in agreement with Fig.\,\ref{fig:freqspec}, in which, at speeds lower than critical speed the real part of the first eigenvalue is zero. 
 However, at super-critical speed ${\bar{v}}_\text{sup}$, both the energy rate and displacement amplitude show rapid growth. At this speed, as shown in Fig.\,\ref{fig:freqspec}, $\text{Re}(\omega_1^\text{sup})$ is positive and $\text{Im}(\omega_1^\text{sup})$ 
 vanishes, so that the displacement ${\bar{y}}_\text{sup}$ grows exponentially with time at the rate $\omega_1^\text{sup}=\text{Re}(\omega_1^\text{sup})=0.068$ as shown in the inset of Fig.\,\ref{fig:displenergyH}. Also note that the growth is linear and small for a while before taking off exponentially. With further increase in speed this exponent becomes larger and the growth in amplitude is even faster.  
    Gravity, as discussed previously, only worsens the stability, as shown in  Fig.\,\ref{fig:displenergyV}. Thus, in Fig.\,\ref{fig:displenergyV}, the growth in amplitude for super-critical speed appears earlier and faster in time as compared to the horizontally traveling cable. We observe from the inset of Fig.\,\ref{fig:displenergyV}, in contrast to the inset of Fig.\,\ref{fig:displenergyH}, that  during vertical travel the initial growth is fairly rapid, which causes the amplitude to shoot up before growing exponentially at the same rate $\omega_1^\text{sup}=0.068$. From this we may infer that the instability in operation  appears earlier in an inclined traveling cable than in a horizontal cable. Thus, the results of energetics and transient analysis support those found from modal analysis. 

The super-critical energy rate $\dot{\bar{E}}_\text{sup}$ grows monotonically  for a horizontally traveling cable (Fig.\,\ref{fig:displenergyH}), whereas the growth is non-monotonous in case of vertically traveling cable (Fig.\,\ref{fig:displenergyV}). This non-monotonous growth in the latter case is due to the second term in \eqref{eq:EdotV}, which is a contribution of cable's weight that is absent during horizontal travel. Unlike the first term in \eqref{eq:EdotV}, which is calculated only at the boundaries, the second term is obtained after integration over the cable's length. Therefore, at super-critical speed, the purely exponential growth of ${\bar{y}}_\text{sup}$ causes $\dot{\bar{E}}_\text{sup}$ to grow non-linearly for both horizontally and vertically traveling cables, while the averaged quantity introduces oscillations in $\dot{\bar{E}}_\text{sup}$ for the vertically traveling cables.

\section{Conclusions} \label{sec-conclusion}
In this paper, we studied the vibration analysis of a tensioned, heavy cable traveling inclined to gravity with a constant speed. The cable was modeled as a beam resisting tension, shear and bending. Modal analysis was performed for both the Rayleigh and the Euler-Bernoulli beam models. We identified the critical values of those parameters at which instability occurs; namely  inclination, speed of travel, end-tension, slenderness ratio and bending rigidity. Stability curves, or boundaries, were obtained in the space of these parameters. The transient and energy analyses on the Euler-Bernoulli model were used to verify these boundaries. As a special case, highly flexible cables with zero bending rigidity were modeled as strings. 

The instability in modal analysis was identified with the critical value of parameter at which the real part of the smallest eigenvalue first becomes positive. For the Euler-Bernoulli beam the critical speed $\bar{v}_\text{crit}$  coincides with the speed $\bar{v}_\text{bif}$ corresponding to the bifurcation of the real part of eigenvalue  and simultaneous vanishing of the associated imaginary part. Physically, this phenomenon corresponds to exponential growing of the amplitude of the displacement $\bar{y}_\text{sup}$ at a rate $\omega_1^\text{sup} = \text{Re}\left( \omega_1^\text{sup} \right)$, while the rate of energy influx also appears to rise.  In contrast to Euler-Bernoulli beam, $\bar{v}_\text{crit}$ and $\bar{v}_\text{bif}$  are different for a Rayleigh beam. 

Our observations are as follows. Firstly, the applied end-tension enhances the stability of a traveling cable system, while the action of gravity, due to the cable's inclination, lowers it. This is observed regardless of whether   the traveling cable is modeled as a beam or as a string.  Secondly, slenderness ratio improves the stability of the system. However, the effect of slenderness is more clearly observed when a traveling cable is modeled as a Rayleigh beam rather than an Euler-Bernoulli beam. When slenderness ratios are high both the models give close results. Thirdly, while an inclined traveling Rayleigh beam shows instability for all travel speeds, i.e. $\bar{v}_\text{crit}=0$ and $\bar{v}_\text{bif}>\bar{v}_\text{crit}$,  the effects of inclination on the stability of Euler-Bernoulli beam are negligible. However, the inherent instability of Rayleigh beam can be removed by inclusion of damping. When damping is included, $\bar{v}_\text{bif}<\bar{v}_\text{crit}$ and the Rayleigh beam undergoes underdamped oscillations at sub-bifurcation travel speeds. In Euler-Bernoulli beams the damping further enhances the stability. Lastly, the time rate of change of energy $\dot{\bar{E}}$ of the traveling cable is not constant, exhibiting a non-conservative behaviour. This behaviour doesn't necessarily imply instability at all speeds of travel. At sub-critical speeds, $\dot{\bar{E}}_\text{sub}$ is oscillatory and remains bounded for extended period of time. Whereas, at super-critical speeds $\dot{\bar{E}}_\text{sup}$ grows monotonically for horizontally traveling cables and non-monotonically for inclined traveling cables. Both $\dot{\bar{E}}_\text{sup}$ and $\bar{y}_\text{sup}$ grow rapidly at higher inclinations, thus, confirming the destabilizing effect caused by the action of gravity. 


\appendix
\section{Convergence of solution}
\label{appendix-convergence}
\pgfplotstableread[col sep = comma]{betaV5_0.txt}\bvert
\pgfplotstableread[col sep = comma]{betaV5_90.txt}\bhoriz
\pgfplotstableread[col sep = comma]{CriticalV_0.txt}\Criticalvert
\pgfplotstableread[col sep = comma]{CriticalV_90.txt}\Criticalhoriz
The converged values of critical speed for different inclinations are
found by progressively increasing the number $N$ of sine functions until the relative error between two successive computations
becomes lower than $0.001\%$. For the horizontally traveling cable, the computed
values of critical speed from beam and string models are compared with
those obtained by employing Green's functions in \cite{wickert1990classical}. Additionally, we solved (\ref{eq:mateqn}) numerically utilizing 2-node Hermitian finite elements (FE). The converged solution from this procedure is in excellent agreement with that found through Galerkin projection with Fourier sine bases. Figure \ref{fig:subcriteigerr}(\textbf{a}) plots the relative error 
\[\varepsilon_\text{crit}^v = \left| 1 - \dfrac{v_\text{crit}^{N} }{ v_\text{crit}^\text{ref} }  \right| \cdot 100 \]
in calculating the critical speed for an Euler-Bernoulli beam at different values of $N$. Similarly, Fig.\,\ref{fig:subcriteigerr}(\textbf{b}) plots the relative error
\[\varepsilon_\text{sub}^\omega = \left| 1 - \dfrac{ {\omega_1^N  } }{ { \omega_1^\text{ref}  } } \right| \cdot 100 \]
in calculation of the first eigenvalue at a sub-critical speed. 
For horizontally traveling beams the closed form results of  \cite{wickert1990classical} are taken as reference. 
 We conclude that both FE solution and Galerkin projection with Fourier sine bases are accurate when $N \geq 9$.
%
\begin{figure}[h!]
  \begin{center}
%
\pgfplotsset{
  every axis plot post/.append style={
   black,
   every mark/.append style={scale=1,solid}
   }}
\begin{tikzpicture}
\begin{groupplot}[
  cycle list name=black white,
     group style={
        group name=my fancy plots,
        group size=1 by 2,
        xticklabels at=edge bottom,
        vertical sep=10pt
    },
    width=8.5cm,
]
\nextgroupplot[
    ylabel={Error \%, $\varepsilon_\text{crit}^v$},
    xmin = 1,
    xmax = 13,
    ymin = -0.1,
    ymax = 1 ,
    height=5.0cm,
    axis x line=bottom,
    axis y line=left,
    table/x index = 0 ,
    legend columns=3,
     legend style={fill=none,column sep=0.2ex,at={(0.5,1.4)},anchor=north,cells={anchor=west}}, 
               ]
\addlegendimage{empty legend}
  \addlegendentry{For $\varphi=0$: }               
\addplot [mark=square*,mark options={fill=white}] table[ y expr=abs(10.48-\thisrow{Sine})/0.1048,col sep=comma] {\Criticalhoriz};  \addlegendentry{sine \quad}
\addplot [mark=square*,mark options={fill=white},loosely dotted,very thick] table[ y expr=abs(10.48-\thisrow{FEM})/0.1048,col sep=comma] {\Criticalhoriz};  \addlegendentry{FE}
\addlegendimage{empty legend}
  \addlegendentry{For $\varphi=\pi/2$: }               
\addplot [mark=*,mark options={fill=white}] table[ y expr=abs(10.23-\thisrow{Sine})/0.1023,col sep=comma] {\Criticalvert};  \addlegendentry{sine \quad}
\addplot [mark=*,mark options={fill=white},loosely dotted,very thick] table[ y expr=abs(10.23-\thisrow{FEM})/0.1023,col sep=comma] {\Criticalvert};  \addlegendentry{FE}
\node at (axis cs:13,1) [anchor=north east,fill=none] {(\textbf{a})};
\nextgroupplot[xlabel={$N$},
    ylabel={Error \%, $\varepsilon_\text{sub}^\omega$},
    xmin = 1,
    xmax = 13,
    ymin = -0.1,
    axis x line=bottom,
    axis y line=left,
    height=5.0cm,
    table/x index = 0 ,
    ]
\addplot [mark=*,mark options={fill=white}] table[ y expr=abs(25.59-\thisrow{Sine})/0.2559,col sep=comma] {\bvert}; 
\addplot [mark=*,mark options={fill=white},loosely dotted,very thick] table[ y expr=abs(25.59-\thisrow{FEM})/0.2559,col sep=comma] {\bvert}; 
\addplot [mark=square*,mark options={fill=white},mark options={fill=white}] table[ y expr=abs(26.47-\thisrow{Sine})/0.2647,col sep=comma] {\bhoriz};  
\addplot [mark=square*,mark options={fill=white},loosely dotted,very thick] table[ y expr=abs(26.47-\thisrow{FEM})/0.2647,col sep=comma] {\bhoriz};  
\node at (axis cs:13,0.7) [anchor=north east,fill=none] {(\textbf{b})};
\end{groupplot}
\end{tikzpicture}
%
  \end{center}
    \caption{Plot (\textbf{a}) compares the errors in critical speed $v_\text{crit}^N$ obtained from the $N$-term (Fourier) sine approximation and the $N$-element FE solution at various inclinations. In each case $\mu=10$. For $\varphi = 0$ (horizontally traveling cable) the critical speed  $v_{\text{crit}}^{\text{ref}} = 10.48$, as found by Wickert and Mote, 1990. For $\varphi=\pi/2$, the critical velocity calculated using $N=399$ terms is taken as the reference. Plot (\textbf{b}) compares errors in the calculation of the first eigenvalue $\omega_1$ at sub-critical speed $v_\text{sub}^\text{ref} = 5$. } \label{fig:subcriteigerr}
\end{figure}
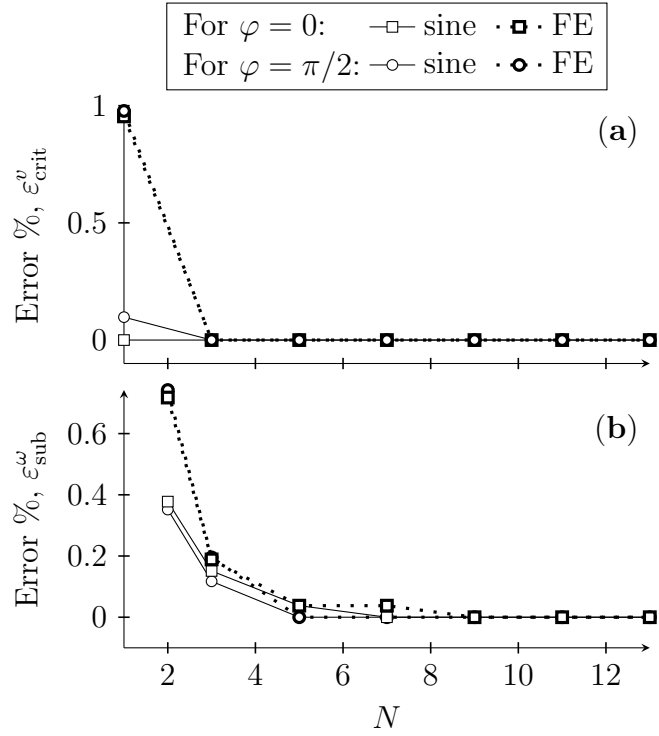
%

\singlespacing
\providecommand{\noopsort}[1]{}\providecommand{\singleletter}[1]{#1}%

\end{document}

%% file: beam.pdf_tex
\begingroup%
  \makeatletter%
  \providecommand\color[2][]{%
    \errmessage{(Inkscape) Color is used for the text in Inkscape, but the package 'color.sty' is not loaded}%
    \renewcommand\color[2][]{}%
  }%
  \providecommand\transparent[1]{%
    \errmessage{(Inkscape) Transparency is used (non-zero) for the text in Inkscape, but the package 'transparent.sty' is not loaded}%
    \renewcommand\transparent[1]{}%
  }%
  \providecommand\rotatebox[2]{#2}%
  \ifx\svgwidth\undefined%
    \setlength{\unitlength}{242bp}
    \ifx\svgscale\undefined%
      \relax%
    \else%
      \setlength{\unitlength}{\unitlength * \real{\svgscale}}%
    \fi%
  \else%
    \setlength{\unitlength}{\svgwidth}%
  \fi%
  \global\let\svgwidth\undefined%
  \global\let\svgscale\undefined%
  \makeatother%
  \begin{picture}(1,1)
    \put(0,0){\includegraphics[width=\unitlength,page=1]{beam.pdf}}%
    \put(0.09222592,0.8359757){\color[rgb]{0,0,0}\makebox(0,0)[rb]{\smash{Cable}}}%
    \put(0.09222592,0.8009757){\color[rgb]{0,0,0}\makebox(0,0)[rb]{\smash{Element}}}%
    \put(0.07643537,0.69883655){\color[rgb]{0,0,0}\makebox(0,0)[lb]{\smash{$M$}}}%
    \put(0.08371492,0.54688515){\color[rgb]{0,0,0}\makebox(0,0)[lb]{\smash{$T$}}}%
    \put(0.2274862,0.52596585){\color[rgb]{0,0,0}\makebox(0,0)[lb]{\smash{$V$}}}%
    \put(0.40,0.97544098){\color[rgb]{0,0,0}\makebox(0,0)[lb]{\smash{$T+\partial_x T dx$}}}%
    \put(0.41,0.90){\color[rgb]{0,0,0}\makebox(0,0)[lb]{\smash{$M+\partial_x M dx$}}}
    \put(0.04,0.90540344){\color[rgb]{0,0,0}\makebox(0,0)[lb]{\smash{$V+\partial_x V dx$}}}%
    \put(0.59518458,0.77){\color[rgb]{0,0,0}\makebox(0,0)[lb]{\smash{$v$}}}%
    \put(0.82513044,0.76069405){\color[rgb]{0,0,0}\makebox(0,0)[lb]{\smash{Cable at}}}
    \put(0.80513044,0.71669405){\color[rgb]{0,0,0}\makebox(0,0)[lb]{\smash{equilibrium}}} 
    \put(0.72005528,0.64015924){\color[rgb]{0,0,0}\makebox(0,0)[lb]{\smash{$y(x,t)$}}}%
    \put(0.62194257,0.49953052){\color[rgb]{0,0,0}\makebox(0,0)[lb]{\smash{Cable at}}}%
    \put(0.63594257,0.45953052){\color[rgb]{0,0,0}\makebox(0,0)[lb]{\smash{time $t$}}}%
    \put(0.42767405,0.2893614){\color[rgb]{0,0,0}\makebox(0,0)[lb]{\smash{$F$}}}%
    \put(0.3294,0.49682876){\color[rgb]{0,0,0}\makebox(0,0)[lb]{\smash{$g$}}}%
    \put(0.00,0.31847964){\color[rgb]{0,0,0}\makebox(0,0)[lb]{\smash{$Y$}}}%
    \put(0.22020663,0.35305754){\color[rgb]{0,0,0}\makebox(0,0)[lb]{\smash{$X$}}}%
    \put(0.075,0.09545311){\color[rgb]{0,0,0}\makebox(0,0)[lb]{\smash{$\varphi$}}}%
    \put(0.68,0.32521885){\color[rgb]{0,0,0}\makebox(0,0)[lb]{\smash{$L$}}}%
    \put(0.24932487,0.65153825){\color[rgb]{0,0,0}\makebox(0,0)[lb]{\smash{$wdx$}}}%
    \put(0.33485968,0.77891171){\color[rgb]{0,0,0}\makebox(0,0)[lb]{\smash{$Fdx$}}}%
    \put(0.2424862,0.75569405){\color[rgb]{0,0,0}\makebox(0,0)[lb]{\smash{$dx$}}}%
  \end{picture}%
\endgroup%

%% file: freqspec.pdf_tex
\begingroup%
  \makeatletter%
  \providecommand\color[2][]{%
    \errmessage{(Inkscape) Color is used for the text in Inkscape, but the package 'color.sty' is not loaded}%
    \renewcommand\color[2][]{}%
  }%
  \providecommand\transparent[1]{%
    \errmessage{(Inkscape) Transparency is used (non-zero) for the text in Inkscape, but the package 'transparent.sty' is not loaded}%
    \renewcommand\transparent[1]{}%
  }%
  \providecommand\rotatebox[2]{#2}%
  \ifx\svgwidth\undefined%
    \setlength{\unitlength}{242.28014404bp}%
    \ifx\svgscale\undefined%
      \relax%
    \else%
      \setlength{\unitlength}{\unitlength * \real{\svgscale}}%
    \fi%
  \else%
    \setlength{\unitlength}{\svgwidth}%
  \fi%
  \global\let\svgwidth\undefined%
  \global\let\svgscale\undefined%
  \makeatother%
  \begin{picture}(1,0.72009755)%
    \put(0,0){\includegraphics[width=\unitlength]{freqspec.pdf}}
    \put(0.0655005,0.23143856){\color[rgb]{0,0,0}\makebox(0,0)[lb]{\smash{$0$}}}%
    \put(0.0549884,0.36351706){\color[rgb]{0,0,0}\makebox(0,0)[lb]{\smash{$10$}}}%
    \put(0.02260723,0.09576003){\color[rgb]{0,0,0}\makebox(0,0)[lb]{\smash{$-10$}}}%
    \put(0.0655005,0.55503084){\color[rgb]{0,0,0}\makebox(0,0)[lb]{\smash{$0$}}}%
    \put(0.0549884,0.6937133){\color[rgb]{0,0,0}\makebox(0,0)[lb]{\smash{$10$}}}%
    \put(0.02260723,0.42955629){\color[rgb]{0,0,0}\makebox(0,0)[lb]{\smash{$-10$}}}%
    \put(0.1146389,0.05036244){\color[rgb]{0,0,0}\makebox(0,0)[lb]{\smash{$0$}}}%
    \put(0.24671742,0.05036244){\color[rgb]{0,0,0}\makebox(0,0)[lb]{\smash{$1$}}}%
    \put(0.37879593,0.05036244){\color[rgb]{0,0,0}\makebox(0,0)[lb]{\smash{$2$}}}%
    \put(0.51747836,0.05036244){\color[rgb]{0,0,0}\makebox(0,0)[lb]{\smash{$3$}}}%
    \put(0.64955687,0.05036244){\color[rgb]{0,0,0}\makebox(0,0)[lb]{\smash{$4$}}}%
    \put(0.78823939,0.05036244){\color[rgb]{0,0,0}\makebox(0,0)[lb]{\smash{$5$}}}%
    \put(0.92031796,0.05036244){\color[rgb]{0,0,0}\makebox(0,0)[lb]{\smash{$6$}}}%
    \put(0.36546351,0.00243417){\color[rgb]{0,0,0}\makebox(0,0)[lb]{\smash{Nondimensional speed, $\bar{v}$}}}%
    \put(0.01254359,0.18324151){\color[rgb]{0,0,0}\rotatebox{90}{\makebox(0,0)[lb]{\smash{$\text{Im}(\omega)$}}}}%
    \put(0.01254359,0.51961363){\color[rgb]{0,0,0}\rotatebox{90}{\makebox(0,0)[lb]{\smash{$\text{Re}(\omega)$}}}}%
    \put(0.283183,0.3812959){\color[rgb]{0,0,0}\makebox(0,0)[lb]{\smash{$\varphi=\pi/2$}}}%
    \put(0.423183,0.3412959){\color[rgb]{0,0,0}\makebox(0,0)[lb]{\smash{$\pi/4$}}}%
    \put(0.535,0.31030374){\color[rgb]{0,0,0}\makebox(0,0)[lb]{\smash{$0$}}}%
    \put(0.247,0.51638542){\color[rgb]{0,0,0}\makebox(0,0)[lb]{\smash{$\varphi=\pi/2$}}}%
    \put(0.39558634,0.48038542){\color[rgb]{0,0,0}\makebox(0,0)[lb]{\smash{$\pi/4$}}}%
    \put(0.54,0.45038542){\color[rgb]{0,0,0}\makebox(0,0)[lb]{\smash{$0$}}}%
    \put(0.35001271,0.62786911){\color[rgb]{0,0,0}\makebox(0,0)[lb]{\smash{$2.387$}}}%
    \put(0.41000000,0.67732635){\color[rgb]{0,0,0}\makebox(0,0)[lb]{\smash{$2.692$}}}%
    \put(0.50000000,0.72000000){\color[rgb]{0,0,0}\makebox(0,0)[lb]{\smash{$3.2969$}}}%
    %
    \put(0.58,0.38){\color[rgb]{0,0,0}\makebox(0,0)[lb]{\smash{higher mode}}}%
    \put(0.59,0.33){\color[rgb]{0,0,0}\makebox(0,0)[lb]{\smash{mode mixing}}}%
  \end{picture}%
\endgroup%

%% file: freqspecC_c0.pdf_tex
\begingroup%
  \makeatletter%
  \providecommand\color[2][]{%
    \errmessage{(Inkscape) Color is used for the text in Inkscape, but the package 'color.sty' is not loaded}%
    \renewcommand\color[2][]{}%
  }%
  \providecommand\transparent[1]{%
    \errmessage{(Inkscape) Transparency is used (non-zero) for the text in Inkscape, but the package 'transparent.sty' is not loaded}%
    \renewcommand\transparent[1]{}%
  }%
  \providecommand\rotatebox[2]{#2}%
  \ifx\svgwidth\undefined%
    \setlength{\unitlength}{250bp}
    \ifx\svgscale\undefined%
      \relax%
    \else%
      \setlength{\unitlength}{\unitlength * \real{\svgscale}}%
    \fi%
  \else%
    \setlength{\unitlength}{\svgwidth}%
  \fi%
  \global\let\svgwidth\undefined%
  \global\let\svgscale\undefined%
  \makeatother%
  \begin{picture}(1,1)
    \put(0,0){\includegraphics[width=\unitlength,page=1]{freqspecC_c0.pdf}}%
    \put(-0.052,0.7529344){\color[rgb]{0,0,0}\makebox(0,0)[lb]{\smash{$10$}}}%
    \put(-0.032,0.66348658){\color[rgb]{0,0,0}\makebox(0,0)[lb]{\smash{$5$}}}%
    \put(-0.032,0.57810491){\color[rgb]{0,0,0}\makebox(0,0)[lb]{\smash{$0$}}}%
    \put(-0.032,0.48865732){\color[rgb]{0,0,0}\makebox(0,0)[lb]{\smash{$5$}}}%
    \put(-0.052,0.40734135){\color[rgb]{0,0,0}\makebox(0,0)[lb]{\smash{$10$}}}%
    \put(-0.052,0.35042051){\color[rgb]{0,0,0}\makebox(0,0)[lb]{\smash{$10$}}}%
    \put(-0.032,0.26097278){\color[rgb]{0,0,0}\makebox(0,0)[lb]{\smash{$5$}}}%
    \put(-0.032,0.17559116){\color[rgb]{0,0,0}\makebox(0,0)[lb]{\smash{$0$}}}%
    \put(-0.032,0.08614365){\color[rgb]{0,0,0}\makebox(0,0)[lb]{\smash{$5$}}}%
    \put(-0.052,0.00482768){\color[rgb]{0,0,0}\makebox(0,0)[lb]{\smash{$10$}}}%
    \put(-0.08,0.41140716){\color[rgb]{0,0,0}\makebox(0,0)[lb]{\smash{$-$}}}%
    \put(-0.07,0.48865728){\color[rgb]{0,0,0}\makebox(0,0)[lb]{\smash{$-$}}}%
    \put(-0.08,0.00889352){\color[rgb]{0,0,0}\makebox(0,0)[lb]{\smash{$-$}}}%
    \put(-0.07,0.08614362){\color[rgb]{0,0,0}\makebox(0,0)[lb]{\smash{$-$}}}%
    \put(0.11,0.72548993){\color[rgb]{0,0,0}\makebox(0,0)[lb]{\smash{$0$}}}%
    \put(0.085,0.76208215){\color[rgb]{0,0,0}\makebox(0,0)[lb]{\smash{$0.01$}}}%
    \put(0.085,0.6888978){\color[rgb]{0,0,0}\makebox(0,0)[lb]{\smash{$0.01$}}}%
    \put(0.054,0.69321758){\color[rgb]{0,0,0}\makebox(0,0)[lb]{\smash{$-$}}}%
    \put(0.00117909,-0.03738366){\color[rgb]{0,0,0}\makebox(0,0)[lb]{\smash{$0$}}}%
    \put(0.25020902,-0.03737517){\color[rgb]{0,0,0}\makebox(0,0)[lb]{\smash{$1$}}}%
    \put(0.5005095,-0.03736704){\color[rgb]{0,0,0}\makebox(0,0)[lb]{\smash{$2$}}}%
    \put(0.75589219,-0.03733451){\color[rgb]{0,0,0}\makebox(0,0)[lb]{\smash{$3$}}}%
    \put(0.165,0.64){\color[rgb]{0,0,0}\makebox(0,0)[lb]{\smash{$0$}}}%
    \put(0.34,0.64){\color[rgb]{0,0,0}\makebox(0,0)[lb]{\smash{$1$}}}%
    \put(0.52,0.64){\color[rgb]{0,0,0}\makebox(0,0)[lb]{\smash{$2$}}}%
    \put(0.23,0.8190601){\color[rgb]{0,0,0}\makebox(0,0)[lb]{\smash{$\bar{v}_\text{crit}=0$}}}%
    \put(0.69,0.75){\color[rgb]{0,0,0}\makebox(0,0)[lb]{\smash{bifurcation}}}%
    \put(0.56,0.35908047){\color[rgb]{0,0,0}\makebox(0,0)[lb]{\smash{$\bar{v}_\text{bif}=2.671$}}}%
    \put(-0.06552377,0.56754749){\color[rgb]{0,0,0}\rotatebox{90}{\makebox(0,0)[lb]{\smash{Re($\omega$)}}}}%
    \put(-0.0646293,0.16505312){\color[rgb]{0,0,0}\rotatebox{90}{\makebox(0,0)[lb]{\smash{Im($\omega$)}}}}%
    \put(0.26,-0.08){\color[rgb]{0,0,0}\makebox(0,0)[lb]{\smash{Nondimensional speed, $\bar{v}$}}}%
  \end{picture}%
\endgroup%

%% file: freqspecC_c0510.pdf_tex
\begingroup%
  \makeatletter%
  \providecommand\color[2][]{%
    \errmessage{(Inkscape) Color is used for the text in Inkscape, but the package 'color.sty' is not loaded}%
    \renewcommand\color[2][]{}%
  }%
  \providecommand\transparent[1]{%
    \errmessage{(Inkscape) Transparency is used (non-zero) for the text in Inkscape, but the package 'transparent.sty' is not loaded}%
    \renewcommand\transparent[1]{}%
  }%
  \providecommand\rotatebox[2]{#2}%
  \ifx\svgwidth\undefined%
    \setlength{\unitlength}{250bp}
    \ifx\svgscale\undefined%
      \relax%
    \else%
      \setlength{\unitlength}{\unitlength * \real{\svgscale}}%
    \fi%
  \else%
    \setlength{\unitlength}{\svgwidth}%
  \fi%
  \global\let\svgwidth\undefined%
  \global\let\svgscale\undefined%
  \makeatother%
  \begin{picture}(1,1)
    \put(0,0){\includegraphics[width=\unitlength,page=1]{freqspecC_c0510.pdf}}%
    \put(-0.038,0.899622){\color[rgb]{0,0,0}\makebox(0,0)[lb]{\smash{$0$}}}%
    \put(-0.038,0.98281285){\color[rgb]{0,0,0}\makebox(0,0)[lb]{\smash{$2$}}}%
    \put(-0.038,1.06637313){\color[rgb]{0,0,0}\makebox(0,0)[lb]{\smash{$4$}}}%
    \put(-0.038,1.1398906){\color[rgb]{0,0,0}\makebox(0,0)[lb]{\smash{$6$}}}%
    \put(-0.038,0.65391877){\color[rgb]{0,0,0}\makebox(0,0)[lb]{\smash{$6$}}}%
    \put(-0.038,0.73710962){\color[rgb]{0,0,0}\makebox(0,0)[lb]{\smash{$4$}}}%
    \put(-0.038,0.82066991){\color[rgb]{0,0,0}\makebox(0,0)[lb]{\smash{$2$}}}%
    \put(-0.038,0.27665792){\color[rgb]{0,0,0}\makebox(0,0)[lb]{\smash{$0$}}}%
    \put(-0.038,0.35984877){\color[rgb]{0,0,0}\makebox(0,0)[lb]{\smash{$2$}}}%
    \put(-0.038,0.44340903){\color[rgb]{0,0,0}\makebox(0,0)[lb]{\smash{$4$}}}%
    \put(-0.038,0.51692655){\color[rgb]{0,0,0}\makebox(0,0)[lb]{\smash{$6$}}}%
    \put(-0.038,0.03095471){\color[rgb]{0,0,0}\makebox(0,0)[lb]{\smash{$6$}}}%
    \put(-0.038,0.11414556){\color[rgb]{0,0,0}\makebox(0,0)[lb]{\smash{$4$}}}%
    \put(-0.038,0.19770585){\color[rgb]{0,0,0}\makebox(0,0)[lb]{\smash{$2$}}}%
    \put(-0.0747266,0.82684341){\color[rgb]{0,0,0}\makebox(0,0)[lb]{\smash{$-$}}}%
    \put(-0.0747266,0.74041468){\color[rgb]{0,0,0}\makebox(0,0)[lb]{\smash{$-$}}}%
    \put(-0.0747266,0.66015928){\color[rgb]{0,0,0}\makebox(0,0)[lb]{\smash{$-$}}}%
    \put(-0.0747266,0.20332163){\color[rgb]{0,0,0}\makebox(0,0)[lb]{\smash{$-$}}}%
    \put(-0.0747266,0.11689288){\color[rgb]{0,0,0}\makebox(0,0)[lb]{\smash{$-$}}}%
    \put(-0.0747266,0.0366375){\color[rgb]{0,0,0}\makebox(0,0)[lb]{\smash{$-$}}}%
    \put(-0.00129642,-0.03218657){\color[rgb]{0,0,0}\makebox(0,0)[lb]{\smash{$2$}}}%
    \put(0.19217068,-0.03218657){\color[rgb]{0,0,0}\makebox(0,0)[lb]{\smash{$2.2$}}}%
    \put(0.3895071,-0.03287801){\color[rgb]{0,0,0}\makebox(0,0)[lb]{\smash{$2.4$}}}%
    \put(0.5833437,-0.03221068){\color[rgb]{0,0,0}\makebox(0,0)[lb]{\smash{$2.6$}}}%
    \put(0.78068012,-0.03221656){\color[rgb]{0,0,0}\makebox(0,0)[lb]{\smash{$2.8$}}}%
    \put(0.26,-0.08){\color[rgb]{0,0,0}\makebox(0,0)[lb]{\smash{Nondimensional speed, $\bar{v}$}}}%
    \put(-0.08,0.90410638){\color[rgb]{0,0,0}\rotatebox{90}{\makebox(0,0)[lb]{\smash{Re($\omega$)}}}}%
    \put(-0.08,0.28114231){\color[rgb]{0,0,0}\rotatebox{90}{\makebox(0,0)[lb]{\smash{Im($\omega$)}}}}%
    \put(0.74,1.17){\color[rgb]{0,0,0}\makebox(0,0)[lb]{\smash{$\bar{v}_\text{crit}=2.759$}}}%
    \put(0.78,1.12){\color[rgb]{0,0,0}\makebox(0,0)[lb]{\smash{$(c=10)$}}}%
    \put(0.44,1.07){\color[rgb]{0,0,0}\makebox(0,0)[lb]{\smash{$\bar{v}_\text{crit}=2.692$}}}%
     \put(0.48,1.02){\color[rgb]{0,0,0}\makebox(0,0)[lb]{\smash{$(c=5)$}}}%
    \put(0.04,0.9387173){\color[rgb]{0,0,0}\makebox(0,0)[lb]{\smash{$c=0$}}}%
    \put(0.04,0.84391844){\color[rgb]{0,0,0}\makebox(0,0)[lb]{\smash{$c=5$}}}%
    \put(0.04,0.73796831){\color[rgb]{0,0,0}\makebox(0,0)[lb]{\smash{$c=10$}}}%
    \put(0.26143037,0.47439625){\color[rgb]{0,0,0}\makebox(0,0)[lb]{\smash{$c=0$}}}%
    \put(0.26103037,0.37959739){\color[rgb]{0,0,0}\makebox(0,0)[lb]{\smash{$c=5$}}}%
    \put(0.26103037,0.31234068){\color[rgb]{0,0,0}\makebox(0,0)[lb]{\smash{$c=10$}}}%
    \put(0.11,0.57){\color[rgb]{0,0,0}\makebox(0,0)[lb]{\smash{$2.166$}}}%
    \put(0.35,0.57){\color[rgb]{0,0,0}\makebox(0,0)[lb]{\smash{ $\bar{v}_\text{bif}=$ }}}%
    \put(0.48,0.57){\color[rgb]{0,0,0}\makebox(0,0)[lb]{\smash{$2.555$}}}%
    \put(0.65,0.57){\color[rgb]{0,0,0}\makebox(0,0)[lb]{\smash{$2.671$}}}%
  \end{picture}%
\endgroup%